%% file: assoc-ITA-rev-v6.tex
\documentclass[11 pt, draftclsnofoot, journal, onecolumn]{IEEEtran} 
\usepackage{graphicx,amssymb,amstext,amsmath,color}

\usepackage{mathabx} 

\DeclareGraphicsExtensions{.eps,.pdf,.png,.jpg,.gif,.jpeg,.pstex}

\input{epsf.sty}

\usepackage[normalem]{ulem}
\usepackage{soul}
\usepackage{times, epsf}
\usepackage{amsmath,amssymb}
\usepackage[hyphens]{url}
\usepackage[font = footnotesize]{caption}
\usepackage{ float}
\usepackage{algorithm}
\usepackage{algorithmic}
\usepackage{empheq}
\usepackage{graphicx}
\usepackage{subfig}
\usepackage{cite}
\usepackage{multirow,tabularx}
\usepackage[pdftex]{epsfig}
\input{epsf.sty}

\usepackage{stackrel}
\usepackage{amsfonts}
\input{epsf.sty}

\usepackage{times, epsf}
\usepackage{graphicx}
\usepackage{cite}
\usepackage{multirow,tabularx}
\usepackage{epstopdf}
\usepackage{amsthm}

\newtheorem{theorem}{Theorem}
\newtheorem{proposition}{Proposition}
\newtheorem{definition}{Definition}
\newtheorem{lemma}{Lemma}
\newtheorem{corollary}{Corollary}

\newtheorem{remark}{Remark}

\newcommand{\ba}{\begin{array}}
\newcommand{\ea}{\end{array}}

\def\PARstart#1#2{\begingroup\def\par{\endgraf\endgroup\lineskiplimit=0pt}
    \setbox2=\hbox{\uppercase{#2} }\newdimen\tmpht \tmpht \ht2
    \advance\tmpht by \baselineskip\font\hhuge=cmr10 at \tmpht
    \setbox1=\hbox{{\hhuge #1}}
    \count7=\tmpht \count8=\ht1\divide\count8 by 1000 \divide\count7 by\count8
    \tmpht=.001\tmpht\multiply\tmpht by \count7\font\hhuge=cmr10 at \tmpht
    \setbox1=\hbox{{\hhuge #1}} \noindent \hangindent1.05\wd1
    \hangafter=-2 {\hskip-\hangindent \lower1\ht1\hbox{\raise1.0\ht2\copy1}%
    \kern-0\wd1}\copy2\lineskiplimit=-1000pt}

\newcommand{\bit}{\begin{itemize}}
\newcommand{\eit}{\end{itemize}}

\input{macros}

\def\argmax{\mathop{\rm arg\,max}}

\begin{document}

\title{Optimal User-Cell Association for Massive MIMO Wireless Networks}

\author{{\small Dilip Bethanabhotla$^1$,  Ozgun Y. Bursalioglu$^2$, 
Haralabos C. Papadopoulos$^2$,  and Giuseppe Caire$^1$}
\thanks{$^1$ Department of Electrical Engineering,
University of Southern California, Los Angeles, CA 90089, USA. Email: {\tt bethanab, caire @usc.edu}.}
\thanks{$^2$ Wireless Systems Project, Docomo Innovations Inc, Palo Alto, CA 94304, USA.
Email: {\tt obursalioglu, hpapadopoulos @docomoinnovations.com}}
}
\maketitle

\newpage

\begin{abstract}
The use of a very large number of antennas at the base station sites 
(referred to as {\em Massive MIMO}) is one of the most promising approaches to cope with the 
predicted wireless data traffic explosion.   
Following the current wireless technology trend of moving to higher frequency bands and denser 
cell deployments, a large number of antennas can be implemented within a small form factor 
even in small-cell base stations. 
Envisioned scenarios involve heterogeneous networks (comprised of base stations with different powers, numbers of antennas 
and multiplexing gain capabilities) serving user traffic with often highly non-homogeneous user density. 
A key system optimization problem in such networks consists of associating users to base stations 
such that congestion is avoided and the available wireless infrastructure is efficiently used.

In this paper, we consider the user-cell association problem for a massive MIMO heterogeneous  
network. We formulate the problem as a network utility maximization, where the network utility is a function of the 
users' long-term average rates (per-user throughputs).  
Under a massive-MIMO specific system model, we show that optimizing the activity fractions between user-BS pairs problem 
is a convex problem that can be solved efficiently by centralized sub-gradient algorithms. 
Furthermore, we show that such a solution is {\em physically realizable}, in the sense that there exists 
a scheduling sequence approaching arbitrarily closely the optimal activity fractions. 

We also consider a decentralized user-centric scheme,  where each user has a positive probability to switch cell association if the utility expected from a different base station is higher than the utility achieved from the currently 
associated one. We formulate a non-cooperative association game and show that its pure-strategy Nash equilibria must be close to 
the global optimum of the centralized problem. 
We also show that, under certain technical conditions that we refer to as {\em heavy-loaded network}, if the centralized global optimum consists of a unique association (i.e., no user has positive activity  fraction to more than one base station), then this association is a pure-strategy Nash equilibrium of the corresponding user-centric association game. 
Based on previously known results, we also have that the proposed user-centric decentralized probabilistic scheme converges 
to a pure-strategy Nash equilibrium with probability 1, for the practically relevant cases of proportional fairness 
and max-min fairness utility functions. Hence, our user-centric algorithm is attractive not only for its simplicity  and fully decentralized implementation,
but also because it operates near the system {\em social} optimum.
\end{abstract}


\newpage

\section{Introduction}  \label{sec:intro}

With the proliferation of mobile devices and services, industry predicts that the wireless data traffic is going to increase by two to 
three orders of magnitude within a decade~\cite{Cisco1}. Although the definition of the next generation of systems and standards is at its initial phase, 
it is widely agreed that the next generation of wireless networks,  generally referred to as ``5G'', 
will involve a combination of multiuser MIMO technology, 
cell densification, and heterogeneous architectures based on nested tiers of smaller and smaller cells 
operating at higher and higher frequencies, in order to target traffic hot-spots~\cite{ishii2012novel}. 
These trends have motivated the recent surge of research on massive and dense deployment of base station antennas, 
both in the form of {\em Massive MIMO} schemes, with hundreds of antennas at each cell site \cite{marzetta2010noncooperative, hoydis2011massive, Huh11}, 
and in the form of multi-tier networks of densely deployed small-cells \cite{hoydis2011green, chandrasekhar2008femtocell}. 

Massive MIMO promises dramatic increases in spectral efficiency by transmitting independent data streams simultaneously 
to multiple users sharing the same transmission resource (time-frequency slot). 
The massive MIMO regime \cite{marzetta2010noncooperative,hoydis2011massive, Huh11}
distinguishes itself from classical multiuser MIMO~\cite{quantenna,broadcom}
by the fact that the number of served users is significantly less than the (very large) number of base station antennas. 
Operating in {\em Time-Division Duplexing} (TDD) mode, massive MIMO can provide very large spectral efficiencies, simple per-cell processing, 
and very attractive power efficiency due to the large array gain~\cite{marzetta2010noncooperative}.  
Thanks to the higher and higher carrier frequencies~\cite{rappaport2013mmwave}, it is possible to implement massive MIMO
even in relatively small base stations within a reasonable form factor. Hence, it is envisaged that massive MIMO will not just be applied to large 
tower-mounted base stations, but also used in conjunction with small cells \cite{adhikary2013joint}. 

 The heterogeneous wireless network framework mentioned above may include some of the following features: 1) base stations that may differ significantly by  transmit power, number of antennas, and multiplexing gain (e.g., see \cite{ghosh2012heterogeneous} and references therein); 
2) non-homogeneous user spatial distribution, characterized by  high-density hot-spots separated by less dense regions \cite{3GPP};
3) Due to the large beamforming gain of massive MIMO, a user may be in good SINR conditions with respect to several base stations.
As a consequence, the rationale that has driven for decades the conventional cellular system design and optimization, 
based on symmetric lattice-deployed cells (see for example  \cite{marzetta2010noncooperative, hoydis2011massive, Huh11}) and/or
(roughly) uniform number of users per cell  (e.g., see \cite{dhillon2012modeling} and references therein), 
must be abandoned in favor of more efficient schemes that include user-cell association into the 
optimization problem. 

In conventional technologies, the user-cell association is decided on the basis of the 
so-called {\em Reference Signal Received Power} (RSRP), possibly in combination with 
{\em Reference Signal Received Quality} (RSRQ) (see \cite{3GPP} for details). In short, these are measures of 
the signal strength measured on a load-independent reference beacon signal sent by each base station \cite{andrews2013overview}. 
Such association does not take into account the actual load of base stations, i.e., the number of associated users 
per downlink data stream, and may be arbitrarily suboptimal in a heterogeneous scenario.  
``Biasing'' is a commonly proposed method to cope with cell or user density asymmetries, 
where the RSRP is artificially scaled by a bias term  that depends on the type of base station \cite{ye2012user,andrews2013overview} 
in order to inherently steer users to associate with close  small-cell base stations, thereby ``off-loading''  congested macro-cells. 
Nevertheless, biasing methods are either heuristic or are based on some {\em average} performance metrics, 
where averaging is over the random placement of users and base station according to stochastic geometry models \cite{dhillon2012modeling,andrews2013overview,ye2012user,gupta2013downlink}. Furthermore, biasing attempts to balance user traffic across tiers, but not within each tier. In contrast, here we seek {\em pointwise optimal user-cell association}, i.e., for any given placement of users and base stations.

\subsection{Contributions}

In this paper, we focus on the problem of optimal user-cell association for the downlink
of a heterogeneous wireless network (including the features said above)  with massive MIMO base stations. 
Our problem formulation captures the fact that, in modern data-oriented systems with OFDMA/TDMA scheduling,
not all users are simultaneously served on all the transmission resources (time-frequency slots).  
Hence, what matters is not the user {\em instantaneous} rate or SINR level, achieved 
at any given time-frequency slot, rather the long-term average rate, referred to hereafter as per-user {\em throughput}. 
It is also important to notice that, in realistic network topologies, users have different distances and propagation conditions (path-loss) 
with respect to the base  stations. Hence, maximizing the network spectral efficiency (user sum rate) 
typically yields unacceptable per-user performance, since this may lead to a large number of users located 
in unfavorable positions (e.g., at the cell edges) with near-zero throughput (see  \cite{Huh11,Huh-Tulino-TIT,huh2011multi}). 
Motivated by the above observations, we formulate the system optimization problem as a rigorous Network Utility Maximization (NUM), 
where the fairness criterion is reflected by the choice of the network utility function. Instead of focusing on the per-slot 
instantaneous user rates, our network utility is a function of the user throughputs.  
It should be noticed that fairness across the users is often {\em implicitly} assumed
by considering equal user air-time, as for example in \cite{marzetta2010noncooperative,hoydis2011massive}. 
Since equal air-time is {\em just one of the many possible fairness criteria}, here we take a more systematic 
approach, which includes equal air-time as a special case. 

It is well-known that solving the general joint user-cell association, precoding vectors design, and power allocation problem 
is NP-hard \cite{hong2013distributed,sanjabi2012optimal}.  
Instead, in this paper we heavily exploit the specific system simplification occurring 
in the massive MIMO regime \cite{marzetta2010noncooperative, hoydis2011massive, Huh11}. 
While in general the user instantaneous rates  are functions of the multiuser MIMO precoding scheme, 
of the base station power allocation, and of the MIMO channel matrix realization (e.g., see \cite{caire2010multiuser}), 
in a massive MIMO system the instantaneous rates converge to easily computable {\em deterministic limits} $R_{k,j}$
that depend only on the overall system topology (path gains between base stations user $k$) 
and system configuration (pilot signal allocation, 
transmit power, number of antennas and number of downlink data streams of the base stations), but are independent of
the other users' cell association (see details in Appendix \ref{massiveMIMO}). 
It follows that the massive MIMO regime induces {\em decoupling} and {\em symmetrization} of the user instantaneous rates, 
yielding a dramatic simplification of the NUM problem, which turns out to be convex with respect to the user activity 
fractions, i.e., the fractions $\alpha_{k,j}$  of transmission resources over which user $k$ is served by base station $j$. 
Furthermore, we prove that the solution to this convex problem is {\em physically realizable} in the following sense: 
there exists a feasible schedule consisting of a sequence  of {\it integer scheduling configurations} (see definition in Section \ref{sec:system-model}) 
such that, by time-sharing these configurations, the time-averaged user rates converge to the globally optimal 
throughput vector. 

While our NUM solution is optimal, its implementation as an on-line protocol requires centralized computation and coordination across the base stations. 
This may be undesirable in practice. Then, we also consider a fully decentralized user-centric scheme similar to~\cite{chiang2013ratselection}, 
where each user has a positive probability to switch cell association if the utility expected from a different base station is higher than the utility achieved from the currently associated one. In particular, we formulate a related non-cooperative association game where the users are the players, and the base stations operate according to a local resource allocation rule that determines the users' utility. By studying the KKT conditions of the global optimization problem and comparing them with the conditions under which the best-response strategy of the game makes all users keep their current association, we prove that
the pure-strategy Nash equilibria of such a game must be very close to the global optimum of the centralized problem. Furthermore, 
we prove that, under certain technical conditions that we refer to as {\em heavy-loaded network}, 
if the centralized global optimum consists of a unique association (i.e., no user has positive activity fraction to more than one base station), then 
this association is a pure-strategy Nash equilibrium of the corresponding user-centric association game. 
Based on \cite{chiang2013ratselection}, we also have that the proposed user-centric decentralized 
scheme converges to a Nash equilibrium with probability 1, for the practically relevant cases of proportional fairness (PF) and hard fairness (HF). 
Hence, our user-centric algorithm is attractive not only for its simplicity  and fully decentralized 
implementation, but also because its operates near the system {\em social} optimum.

\subsection{Related work}  \label{sec:work-review}

The need for efficient user-cell association schemes for heterogeneous networks, 
beyond what is currently done for regular/uniform user density deployments, is clearly stated in standard documents such as
\cite{3GPP}. 

The literature on the broad topic of user-cell association is vast, and several approaches  
targeted to different performance metrics and system assumptions have been proposed. 
Providing a compressive coverage of such large body of works would be out of the scope of this paper. 
Therefore, we shall focus only on the works that more directly relate to ours. 

Joint power allocation and user-cell association for the purpose of minimizing the total power subject to target user SINR 
constraints has been widely studied in the framework of CDMA power-controlled networks 
\cite{hanly1995algorithm,son2009dynamic,stanczak2007distributed}.
This approach assumes that all users, on any time-frequency slot, must maintain a certain target instantaneous rate, 
SINR level or, more in general, QoS constraint (see \cite{schubert2006qos}). While this may be relevant for  
CDMA systems, where users continuously transmit (uplink) or receive (downlink), it is not meaningful in the case of OFDMA/TDMA systems
with scheduling, where the instantaneous rate of a user is zero on the slots on which it is not scheduled. 

Several works have considered the problem of joint user-cell association, beamforming vector design and 
power allocation. This problem was shown to be NP-hard \cite{sanjabi2012optimal,razaviyayn2013linear,hong2013distributed} and approximate solution methods have been proposed.
Our work differs from this line of works because the network utility function
used in there is a function of the user instantaneous rates, and the optimization of the powers and beamforming vectors
is based on the instantaneous realization of the channel matrices (see Appendix \ref{massiveMIMO}). The complex channel coefficients
change over time according to the channel coherence time, which may range from a few tens of ms for slowly moving users 
in systems operating in the 2-5 GHz bands, to less than 1 ms for systems operating at mm-waves (e.g., 20-60 GHz \cite{rappaport2013mmwave,adhikary2013joint}). 
It follows that optimizing user-cell association, beamforming vectors and powers on the basis of 
such rapidly varying channel state information is highly impractical, beyond leading to mathematically involved and computationally 
hard problems. 

In \cite{athanasiou2013optimizing}, the problem of user-cell association was considered 
for a particular model of a multi-cell network operating at mm-waves (60 GHz). 
It is assumed that each user and base station are equipped with a steerable antenna and that, for a given association, 
such antennas point perfectly at each other such that they achieve a certain desired gain while rejecting perfectly
the interference from other base stations. This leads to deterministic and decoupled instantaneous user rates $R_{k,j}$. 
In this respect, this system model is similar to ours, where in our case the decoupled and deterministic user instantaneous rates follow from 
the massive MIMO regime (see Section \ref{sec:system-model} and Lemma \ref{massive-lemma}). However, the problem formulation in
\cite{athanasiou2013optimizing} is completely different from ours, since the goal in \cite{athanasiou2013optimizing} consists of 
minimizing the maximum per-base station load, subject to target user throughput  demands. This is a load-balancing problem, while 
here we solve a NUM problem where the user throughputs are not assigned as a constraint, but are the result of 
the optimization.  In addition, \cite{athanasiou2013optimizing} does not apply to a system employing multiuser MIMO at each base station, 
where multiple users can be served simultaneously by a single base station on each slot. 

In \cite{gupta2013downlink}, a multi-tier heterogeneous network with base stations that can possibly use multiuser MIMO
is considered, and the problem of user-cell association is treated from a stochastic geometry viewpoint. 
Base stations in each tier and users are randomly distributed over the system area according to Poisson point processes, 
and an expression for the user SINR averaged over the stochastic base station/user placement is obtained, 
Based on such expression, a biasing scheme is proposed in order to {\em induce} load-balancing between tiers, 
assuming that users connect to the base station with the strongest received signal. While the overall motivation and system
view (a heterogeneous wireless network with multi-antenna base stations) is clearly related to ours, 
the treatment of the problem is clearly completely different. Here, we obtain a pointwise global optimum solution for the users' throughput 
for a given (arbitrary) placement of the users and base stations, while in \cite{gupta2013downlink}, a heuristic biasing scheme is 
obtained on the basis of an SINR performance indicator, with averaging over the ensemble of stochastic placements. 

Our problem formulation is to some extent related to that of \cite{ye2012user}. However, 
\cite{ye2012user} assumes that each base station applies a local PF criterion, and the optimization is given in terms of 
integer (binary 0-1) association variables. The resulting integer-programming problem is
relaxed, and a subgradient method, which can also be seen as an on-line iterative protocol, is proposed.
The same problem is considered in \cite{shen2013downlink}, where Lagrangian duality is used in order to circumvent the 
integer programming problem, and the dual problem is solved via a coordinate descent method, without the need for relaxation. 
Notice that  the problem formulation in \cite{ye2012user,shen2013downlink} applies only 
to base stations serving a single user per slot (no multiuser MIMO), and uniquely to the case where each base station
applies, independently of the others, a local PF policy giving equal air-time to its associated users. 
In contrast, our problem is formulated for a global network utility function (which includes PF at the whole network level)
and our NUM problem is convex in nature, not requiring any convex relaxation.  

On a separate thread, \cite{chiang2013ratselection} proposes a user-centric game-theoretic approach to the association problem, 
which is completely decentralized. The associated randomized algorithm (which can be turned into an on-line protocol) 
is shown to converge to a Nash equilibrium under certain conditions on the per-user utility function. 
The Pareto efficiency of the Nash equilibria is studied, but it is not a priori clear whether such operating points are close to any well-defined 
global system optimality (social welfare).
Our user-centric scheme is closely related to the scheme of \cite{chiang2013ratselection}. However, 
we consider a  more general class of user-centric utility functions reflecting a desired notion of local (per-cell) fairness, 
and we show the non-trivial fact that the corresponding user-centric schemes operate near the system social optimum of the corresponding 
network-wide utility function.

\section{System Model and Problem Definition} \label{sec:system-model}

We consider a system formed by $J$  base stations (BSs) serving $K$ single antenna users, 
distributed over a given area. 
We use $j\in \Jc=\{ 1,2, \ldots, J\}$ and $k\in \Kc=\{ 1,2, \ldots, K\}$ to index base stations 
(BSs) and users respectively.  Each BS schedules transmissions over contiguous time-frequency slots, each comprising a block of OFDM subcarriers 
and symbols.\footnote{For example, 
in LTE \cite{molisch2010wireless}, resource blocks are $7$ OFDM symbols long (corresponding to a duration of $0.5$ ms), and $12$ 
subcarriers wide (corresponding to a bandwidth of$12\times15\mathrm{kHz}=180 \mathrm{kHz}$).} 
We use the commonly accepted and widely used block-fading channel model \cite{caire2010multiuser,marzetta2010noncooperative, hoydis2011massive, Huh11} and distinguish between 
large-scale and small-scale effects. 
The large-scale channel coefficients are functions of the BS-user distance and shadowing. 
The small-scale effects are modeled as Rayleigh fading coefficients that remain constant 
within each slot. We let $M_j$ denote the number of antennas at BS $j$, and $S_j$ denote the number of downlink
data streams that BS $j$ can transmit on any given slot, i.e., $S_j$ is the multiplexing gain of BS $j$ and the ratio $S_j/M_j$ is the corresponding spatial load.  
We assume TDD operation with reciprocity-based channel state estimation \cite{marzetta2010noncooperative,Huh11}. 
Hence, every BS antenna in the vicinity of user $k$ can estimate its downlink channel coefficient to user $k$ from the uplink pilot transmitted 
by user $k$.  This enables the training of large antenna arrays (e.g., $M_j \gg 1$) with training overhead 
proportional to $S_j$.\cite{marzetta2010noncooperative}.

\subsection{Instantaneous rates and user throughput in massive MIMO systems} \label{sec:massive-mimo-rates}

Consider  the rate $R_{k,j}(t)$ that can be reliably transmitted from BS $j$ to user $k$ over a given slot $t$. 
This is referred to as {\em instantaneous user rate}. In general, $R_{k,j}(t)$ depends on both large scale and small scale effects, 
and in particular on the realization on slot $t$ of the $M_j \times S_j$ channel matrix between the antenna array of BS $j$ and the antennas of
the $S_j$ users scheduled on slot $t$ (see Appendix \ref{massiveMIMO}). 
Let $j_k(t)$ denote the index of the BS to which user $k$ is associated at slot time $t$, and let $\Sc_j(t)$ denote the set of 
$S_j$ users scheduled by BS $j$ on slot $t$.  The sequence $\{\jv(t) : t = 1,2,\ldots\}$ with $\jv(t) = (j_1(t), \ldots, j_K(t)) \in \Jc^K$ 
is referred to as an {\em association sequence}. Notice that, in general, the number of users associated to a given BS $j$ at any time $t$ may be different 
from the BS spatial multiplexing gain $S_j$. In particular, if $|\{ k : j_k(t) = j\}| < S_j$, then some downlink data streams are not used, while
if $|\{ k : j_k(t) = j\}| \geq S_j$ then the BS will schedule $S_j$ out of the possible associated users to be served on slot $t$. 
We refer to the users scheduled on a given slot as the {\em active users}, and let  $\Sc_j(t)$
denote the set of active users of BS $j$ at time $t$.
The sequence $\{(\Sc_1(t), \ldots, \Sc_J(t)) : t = 1,2,\ldots\}$ is referred to as an {\em activation sequence}. 

For a given association sequence $\{\jv(t)\}$ and activation sequence $\{(\Sc_1(t), \ldots, \Sc_J(t))\}$,  the throughput of user $k$ is defined as the limit of
the time-averaged scheduled instantaneous rate:\footnote{We denote the indicator function of a condition $\Ac$ as $1\{\Ac\}$.}
\begin{equation} \label{throughput}
r_k = \lim_{T \rightarrow \infty} \frac{1}{T} \sum_{t=1}^T  R_{k,j_k(t)}(t) \times 1\{ k \in \Sc_{j_k(t)}(t) \}, 
\end{equation}
whenever this limit exists (in the sense of convergence in probability \cite{grimmett1992probability}). 
In this work, we restrict to ergodic stationary systems obeying the following assumptions:
\begin{itemize}
\item[A1)] The large-scale channel coefficients are constant in time;
\item[A2)] The small-scale Rayleigh fading coefficients evolve across different slots according to a stationary and ergodic process with given time-frequency  
correlation. 
\item[A3)] The user-cell association policy and scheduling policy at each BS is such that the limit (\ref{throughput}) exists. 
\end{itemize}
While A2) is  a very common assumption in wireless communications \cite{molisch2010wireless}, assumption A1) holds locally, assuming users with 
low mobility with respect to the time scale over which we observe the network. Under A1) and A2), 
assumption A3) is immediately verified by {\em stationary policies}, i.e., policies that 
that determine the user-BS association and sets of active users as a function of the 
the channel coefficients on each slot $t$. 

At this point, we bring in the fundamental system simplification due to the massive MIMO regime with 
per-BS processing  \cite{marzetta2010noncooperative}. 
In our system, the user achievable instantaneous rates $R_{k,j}(t)$ are given by the following  result:

\begin{lemma} \label{massive-lemma}
For given large-scale channel coefficients and assuming that the small-scale Rayleigh fading obeys the mild assumptions in 
 \cite{marzetta2010noncooperative,hoydis2011massive,Huh11}, 
 there exist deterministic quantities $\{R_{k,j}\}$ such that  $R_{k,j}(t) \stackrel{{\rm a.s.}}{\rightarrow} R_{k,j}$, 
 for all $k \in \Kc$ and $j \in \Jc$ as $M_j, S_j \rightarrow \infty$ with fixed spatial load $S_j/M_j = \nu_j \geq 0$. 
 Furthermore, $\{R_{k,j}\}$ are functions of the system parameters but are independent of the 
 user-cell association and of the active user set.
 \end{lemma}
 
 \begin{IEEEproof}
 The proof is a consequence of the large-system analysis based on asymptotic random matrix theory developed in
 \cite{marzetta2010noncooperative,hoydis2011massive,Huh11} for massive MIMO multi-cell systems. For the sake of completeness, 
 in Appendix \ref{massiveMIMO} we provide explicit expressions (taken from  \cite{hoydis2011massive,Huh11}) for the 
 user instantaneous rates $\{R_{k,j}\}$ under various system assumptions.
 \end{IEEEproof}

 As a consequence of Lemma \ref{massive-lemma}, we have
 
 \begin{corollary} \label{throughput-corollary}
 Under the assumptions of Lemma \ref{massive-lemma} and the system assumption A3), the limit in (\ref{throughput})  is given by 
\begin{equation} \label{throughput1}
r_k = \sum_{j \in \Jc} \alpha_{k,j} R_{k,j},  \;\;\;\;\; \forall \; k \in \Kc
\end{equation} 
where $\alpha_{k,j} = \lim_{T \rightarrow \infty} \frac{| \{ t : k \in \Sc_j(t)\}| }{T}$ denotes the limit of the 
fraction of  slots on which user $k$ is served by BS $j$ (activity fraction).
\end{corollary}

\begin{IEEEproof}
It is sufficient to write
\begin{eqnarray} 
\frac{1}{T} \sum_{t=1}^T  R_{k,j_k(t)}(t) \times 1\{ k \in \Sc_{j_k(t)}(t) \} & = & \frac{1}{T} \sum_{j \in \Jc} \sum_{t : j_k(t) = j} R_{k,j}(t) \times 1\{ k \in \Sc_j(t)\} \nonumber \\
& = & \sum_{j \in \Jc} R_{k,j} \frac{1}{T}  \sum_{t : j_k(t) = j} 1\{ k \in \Sc_j(t)\} \label{ziofanale1} \\
& = & \sum_{j \in \Jc} \frac{|\{ t : k \in \Sc_j(t)\}|}{T} R_{k,j} \label{ziofanale2}
\end{eqnarray}
where (\ref{ziofanale1}) follows from Lemma \ref{massive-lemma} and (\ref{ziofanale2}) follows by re-arranging terms and by noticing that
the condition $k \in \Sc_j(t)$ implies that $j_k(t) = j$.
Then, under assumption A3) the limit of the 
fraction of time slots on which user $k$ is active on the downlink of BS $j$, 
$\frac{|\{ t : k \in \Sc_j(t)\}|}{T}$, must exist and it is denoted  by $\alpha_{k,j}$. Thus, taking the limit of (\ref{ziofanale2}) for $T \rightarrow \infty$ 
we find (\ref{throughput1}). 
\end{IEEEproof}

 
It is worthwhile to remark that the (a.s.) convergence in Lemma \ref{massive-lemma} 
is very quick with respect to the $M_j$'s.  In particular,  under mild assumptions on the channel coefficients (in particular, in the assumption of 
Rayleigh fading of Appendix \ref{massiveMIMO}) 
a central limit theorem can be proved such that, for large but finite $M_j$, the actual  
rate can be written as $R_{k,j}(t) = R_{k,j}  + \chi_{k,j}(t)$, where $\chi_{k,j}(t)$ is a Gaussian ``fluctuation''  with mean zero and variance $O(1/M_j^2))$ 
(see for example \cite{couillet2011random} and references therein). 
It follows the asymptotic rate limits $\{R_{k,j}\}$ yield very accurate results even for large but practical values of 
$M_j$ and $S_j$. As a matter of fact, using the asymptotic instantaneous rate limits {\em in lieu} of the corresponding actual quantities 
has become a widely  accepted common practice in massive MIMO system 
analysis \cite{marzetta2010noncooperative,hoydis2011massive,Huh11}. Therefore, we shall use the limiting values $\{R_{k,j}\}$ as a useful and accurate proxy for
the user instantaneous rates. This has the key advantage that the user throughput, of a given network topology, depends on the activity fractions, as seen from (\ref{throughput1}). 
Using this fact, in the next section we shall cast the user-cell association problem as a convex NUM with respect to the variables $\{\alpha_{k,j}\}$.

\subsection{Recasting user-cell association as NUM problem}  \label{sec:system-optimization}

We wish to find the optimal association of users to BSs such that an overall network 
utility function $U(\rv)$ of the user throughputs vector $\rv \eqdef (r_1, \ldots, r_k)$ is maximized. 
We shall choose the network utility function in order to achieve a desired balance between network-wide 
overall performance and user fairness, reflected by the fact that no user should be given zero throughput.\footnote{As a matter of fact, an admission control
scheme at some upper layer decides which users can join the system, such that all admitted users are given positive throughput. In practice, it is meaningless to admit users
and leave them to starve with zero throughput. While we do not treat here admission control, it is meaningful to assume that all users treated by the 
association and scheduling scheme are admitted, and therefore {\em must} be served with some positive throughput.}
Desirable network utility functions $U(\rv)$ are concave and componentwise monotonically increasing, 
such that larger user throughputs yield larger utility, but the shape of the concave function 
imposes the desired notion of fairness. 
In this paper we consider the well-known and widely used family of utility functions defined by~\cite{mo2000fair} 
\begin{equation} \label{utility}
U(\rv) =  \sum_k \phi_\gamma(r_k), 
\end{equation} 
where 
\begin{equation}
\phi_\gamma(x) = \left \{ \begin{array}{ll}
\log x & \mbox{for} \;\; \gamma = 1 \\
\frac{x^{1- \gamma}}{1 - \gamma} & \mbox{for}\;\;  \gamma \neq 1 \end{array} \right. 
\end{equation}
and $\gamma \geq 0$ is a parameter that determines the level of fairness. 
For example,  this family includes PF (for $\gamma = 1$), where 
$U(\rv) = \sum_k \log r_k$, and HF (for $\gamma \rightarrow \infty$), where $U(\rv) = \min_k r_k$. 

In general, we may consider arbitrary restrictions on the possible user-cell associations (e.g., some BSs may have restricted access with respect to certain users). 
Then, we let  $\Jc_k \subseteq \Jc$ denote the set of BSs which can potentially serve user $k$ (of course, the unrestricted access $\Jc_k = \Jc$ is a special case). 
It is important to notice that, though user $k$ is served by a {\it single} BS $j \in \Jc_k$ in any given slot, 
it may be served by {\em different} BSs in $\Jc_k$ on {\em different} slots.  Consequently, a user $k$ can be associated {\it fractionally} 
to multiple BSs in $\Jc_k$. In this case, we have more than a single  activity fraction $\alpha_{k,j} : j \in \Jc_k$ taking positive values. 
We express this notion formally through the following definitions:
\begin{definition}
{\bf Association:} A user $k$ is said to be {\it associated} to the 
set of BSs $\Jc_k^*\subseteq \Jc_k$ if $\alpha_{k,j} >0$ for all $j \in \Jc^*_k$ and 
$\alpha_{k,j}=0$ for all $j \in \Jc_k \setminus \Jc_k^*$. 
\end{definition}
\begin{definition} \label{def:unique-assoc}
{\bf Unique Association}: A user $k$ is said to be {\it uniquely associated} if $|\Jc^*_k| = 1$. 
In this case, we denote by $j_k$ the BS to which user $k$ is uniquely associated, i.e.,  $\Jc^*_k = \{j_k\}$. \hfill$\lozenge$
\end{definition}
Notice that, even though user $k$ is {\it uniquely associated} to BS $j_k$, it is not necessarily {\it served} by BS $j_k$ on all slots.
For example, if $\alpha_{k,j_k} = 0.5$, then BS $j_k$ serves user $k$ only on $50 \%$ of the slots.
\begin{definition}
{\bf Fractional Association}: A user $k$ is said to be {\it fractionally associated} to the set of BSs $\Jc^*_k \subseteq \Jc_k$ 
if $|\Jc^*_k| > 1$. \hfill$\lozenge$
\end{definition}
At this point, the Network Utility Maximization (NUM) problem at hand can be expressed by:
\begin{subequations}\label{OptfixedS}
\begin{align}
 \textrm{maximize} & \;\;\; U(\rv)  \label{maxutil}\\
 \textrm{subject to} & \;\;\; r_k \leq \sum_{j \in \Jc_k} \alpha_{k,j}R_{k,j}, \; \forall \; k \in \Kc \label{equalityconst} \\
&  \;\;\; \sum_{k \in \Kc}\alpha_{k,j} \leq S_j, ~\forall~j \in \Jc \label{matchingconstJ}\\
& \;\;\; \sum_{j \in \Jc}\alpha_{k,j} \leq 1,~\forall~k \in \Kc \label{matchingconstK} \\
& \;\;\; r_k \geq 0, \;\; \alpha_{k,j} \geq 0,~\forall~k \in \Kc,~j \in \Jc, \label{ineq}
\end{align} 
\end{subequations}
where $R_{k,j}$ are the user instantaneous rates given by Lemma \ref{massive-lemma}, 
and where the optimization is with respect to
$\rv$ and $\alphav \eqdef \{\alpha_{k,j}\}$.  An explanation of the constraints (\ref{equalityconst})--(\ref{ineq}) is in order:
\begin{itemize}
\item The constraint (\ref{equalityconst}) follows  from the expression of the user throughput in Corollary \ref{throughput-corollary}, 
when the set of allowed BS is restricted to $\Jc_k$,  
and from the fact that $U(\cdot)$ is componentwise increasing, such that the optimum is always achieved when (\ref{equalityconst}) is satisfied 
with equality for all $k$. 
\item The constraints in~(\ref{matchingconstJ}) reflect the fact that the sum activities of all the users being served by any given BS $j$ 
cannot exceed the number of simultaneous downlink data streams $S_j$ (multiplexing gain constraint). 
\item  The constraint (\ref{matchingconstK}) simply reflects the fact that each user's total activity fraction (over all BSs) 
cannot be more than one (achieving the bound with equality means that a user is served by some BS in every resource block).
\end{itemize}

\begin{remark} \label{remark-gamma1}
An immediate consequence of the network utility function in (\ref{utility}) is that, for $\gamma \geq 1$, the solution of (\ref{OptfixedS}) must associate 
all users, i.e., for all $k \in \Kc$ it must be $|\Jc^*_k| \geq 1$. Otherwise, some user would have zero throughput, yielding $U(\rv) = -\infty$. 
The feature that all users in the system are served with non-zero throughput reflects the notion of fairness built into the NUM problem
and it is very desirable in practice, as we have already remarked. Therefore, from now on, we shall restrict to $\gamma \geq 1$. \hfill$\lozenge$
\end{remark}

We now give some further definitions that will be useful in the sequel.
\begin{definition}  \label{def:feasible-config}
{\bf Feasible Association Configuration}: 
Any set of activity fractions $\{ \alpha_{k,j} \}$ satisfying (\ref{matchingconstJ})-(\ref{ineq}) is said to be a feasible association  configuration.
\hfill$\lozenge$
\end{definition}
\begin{definition} \label{def:integer-config}
{\bf Integer Scheduling Configuration}: A feasible association configuration is said to be an integer 
scheduling configuration if $\alpha_{k,j} \in \{0, 1\}$ for all pairs $(k, j)$.  \hfill$\lozenge$
\end{definition}
In terms of system implementation, it is relevant to consider 
whether a given feasible association configuration can be achieved as a limit, 
for $T \rightarrow \infty$, of the empirical activity fractions resulting from some actual association and activation sequences. 
In particular, we have:
\begin{definition} \label{def:physical-realizability}
{\bf Physically Realizability:} A feasible association configuration $\alphav$ is said to be physically realizable 
if there exist an association sequence $\{\jv(t)\}$ and an activation sequence
$\{(\Sc_1(t), \ldots, \Sc_J(t))\}$ such that $\alpha_{k,j} = \lim_{T \rightarrow \infty} \frac{|\{t : k \in \Sc_j(t)\}|}{T}$ for all $k,j$.  
\hfill$\lozenge$
\end{definition}
Notice that integer scheduling configurations correspond to time-invariant association and activation 
sequences. In fact, in this case the limit of the $(k,j)$-th activity fraction is equal to 1 if user $k$ is permanently 
associated and scheduled (active) on BS $j$, or 0 if it is not associated or it is associated but never scheduled. 
Hence, it is immediate to see that if $\alphav$ is a convex combination of some integer scheduling 
configurations,  then $\alphav$ is physically realizable. Building on this key observation, the following result yields the physical realizability 
(in the sense of Definition \ref{def:physical-realizability}) of the feasible association configurations of the NUM problem (\ref{OptfixedS}). 

\begin{theorem} \label{th-feasibility}
Any feasible association configuration $\alphav$ is physically realizable.
\end{theorem}

\begin{IEEEproof} 
See Appendix \ref{proof:feasibility}.
\end{IEEEproof}

We conclude this section by pointing out some observations about our NUM problem formulation and its solution: 
\begin{itemize}
\item Problem (\ref{OptfixedS}) is {\em convex}. We shall develop in Section \ref{sec:subgradient} an efficient method for its solution 
that also sheds light on the properties of the optimal solution. 
\item Several seemingly similar user-cell association problems have been formulated
assuming that each user is constrained to be associated {\em permanently} to a single BS \cite{ye2012user,bejerano2004fairness,li2008proportional}.
As a consequence,  the resulting optimization is {\em combinatorial}, since it includes 
an additional set of constraints restricting the feasible association configurations to be 
unique associations (see Definition \ref{def:unique-assoc}). 
\item Problem (\ref{OptfixedS}) can be seen as a convex relaxation of the corresponding unique association combinatorial problem. 
Nevertheless,  its solution can be implemented (as a consequence of Theorem \ref{th-feasibility}).
Hence, the resulting optimal utility function value provides a (feasible) upper bound benchmark to any user-cell association scheme 
imposing unique association, for the massive MIMO network with given user instantaneous rates $\{R_{k,j}\}$ and BS multiplexing gains $\{S_j\}$.
\end{itemize}

\section{Centralized Solution}\label{sec:subgradient}

In order to solve the convex program (\ref{OptfixedS}), general purpose numerical solvers like CVX \cite{cvx} or powerful numerical methods such as
the alternating direction methods of multipliers \cite{boyd2011distributed} can be used. 
However, here we focus on solving (\ref{OptfixedS}) by using a direct method based on 
Lagrangian duality. This yields both an efficient numerical method, able to easily handle networks with hundreds of users and tens of base stations, 
and has the non-negligible advantage of illuminating the structure and properties of the solution, which will be used to establish the near-optimality of a decentralized user-centric scheme studied in Section \ref{sec:user centric}.  
We first formulate the dual program of (\ref{OptfixedS}) for a general utility function $U(\cdot)$. 
We then specialize to the class defined in (\ref{utility}) and develop a centralized algorithm to solve the dual program. 
The optimal solution of the dual program is then used to obtain the primal variables $\alphav$.

We form the Lagrangian function for the primal problem (\ref{OptfixedS}) by introducing the dual variables/prices (we use the terms dual variables and prices interchangeably) $\betav = \{\beta_k\}$ for the constraint (\ref{equalityconst}), $\pv = \{p_j\}$ for the constraint (\ref{matchingconstJ}) and 
$\lambdav = \{\lambda_k\}$ for the constraint (\ref{matchingconstK}). Then,  the Lagrangian function takes on the form:
\begin{eqnarray}
 L(\alphav, \rv,\betav,\pv,\lambdav) & = & U(\rv) - \sum_{k}\beta_k \big(r_k-\sum_{j}\alpha_{k,j}R_{k,j}\big)  - \sum_{j}p_j \big( \sum_{k}\alpha_{k,j} - S_j \big ) \notag \\ 
& & - \sum_{k} \lambda_k \big (\sum_{j}\alpha_{k,j} - 1 \big)  \\
& = & U(\rv)  - \sum_{k} \beta_k r_k  + \sum_{j}S_j p_j + \sum_k \lambda_k 
+ \sum_{(k,j)}\alpha_{k,j}(\beta_k R_{k,j}-p_j-\lambda_k). \notag \\
& & \label{lagrangian}
\end{eqnarray}
The dual function is given by the maximum of the Lagrangian over the primal variables $\alphav \geq 0$ and $\rv \geq 0$:
\begin{align}
G(\betav,\pv,\lambdav) = \max_{\alphav, \rv \geq 0}~L(\alphav,\rv,\betav,\pv,\lambdav).
\end{align}
The value $G(\betav,\pv,\lambdav)$, for any set of non-negative dual variables, 
provides an upper bound on the optimal primal objective value. 
The dual program finds the tightest of such upper bounds  by minimizing $G(\betav,\pv,\lambdav)$ over the feasible set 
of dual variables \cite{boyd2009convex}, i.e., it is given by:
\begin{align}
\textrm{minimize}&\;\;\; G(\betav,\pv,\lambdav) \notag \\
\textrm{subject to}&\;\;\; \betav, \pv, \lambdav \geq 0 \notag.
\end{align}
From (\ref{lagrangian}), we observe that $G(\betav,\pv,\lambdav)=\infty$ if $\beta_k R_{k,j}-p_j-\lambda_k > 0$ for some $(k,j)$. In the case when $\beta_k R_{k,j}-p_j-\lambda_k \leq 0$ for all $(k,j)$, it is easy to see that $ L(\alphav, \rv,\betav,\pv,\lambdav)$ is maximized when $\alphav$ is chosen such that the term $\sum_{(k,j)}\alpha_{k,j}(\beta_k R_{k,j}-p_j-\lambda_k)$ in (\ref{lagrangian}) vanishes.
From these observations, we find the dual program equivalent form:
\begin{subequations}
\begin{align}
\textrm{minimize}&\;\;\; \max_{\rv \geq 0} \left \{ U(\rv) - \sum_{k} \beta_k r_k \right \}  + \sum_{j}S_jp_j + \sum_k \lambda_k \label{max-r} \\
\textrm{subject to}&\;\;\; \beta_k R_{k,j} \leq p_j+\lambda_k ~\forall~(k,j)\\
&\;\;\; \betav, \pv, \lambdav \geq 0.
\end{align}
\end{subequations}
We now particularize to the class of network utility functions in (\ref{utility}). 
Thanks to the additive form of the network utility function, the maximization with respect to $\rv$ in (\ref{max-r}) decomposes into the sum (over $k \in \Kc$) of 
individual maximizations of the terms 
\[ \phi_\gamma(r_k) -\beta_k r_k, \;\;\; k \in \Kc. \]
Setting the derivative with respect to $r_k$ to zero, it is immediate to show that the maximum is achieved for
\begin{equation} 
r_k = \beta_k^{-\rho}, \label{mamax}
\end{equation}
where, for future convenience, we define $\rho = 1/\gamma$.  The corresponding (maximum) value is 
$\frac{1}{\rho-1}\beta_k^{1 - \rho}$ for $\gamma \ne 1$, and $-\log \beta_k-1$ for $\gamma=1$.  
We first consider in detail the case $\gamma \ne 1$. 
Using (\ref{mamax}) into (\ref{max-r}), we obtain the dual program in the form:
\begin{subequations}
\begin{align}
\textrm{minimize}&\;\;\;  \sum_{j} S_jp_j + \sum_k \lambda_k + \frac{1}{\rho-1}\sum_k\beta_k^{1- \rho}\\
\textrm{subject to}&\;\;\; \beta_k  \leq \frac{p_j+\lambda_k}{R_{k,j}} ~\forall~(k,j)\\
&\;\;\; \betav, \pv, \lambdav \geq 0. 
\end{align}
\end{subequations}
The minimization over $\betav$ is immediate, and yields
\begin{equation} 
\beta_k = \min_j \left\{\frac{p_j+\lambda_k}{R_{k,j}} \right \}.  \label{mamamax}
\end{equation}
Replacing, we obtain:
\begin{subequations}
\label{target-dual}
\begin{align}
\textrm{minimize}&\;\;\;  \widecheck{G}(\pv, \lambdav)=  \sum_{j}S_jp_j + \sum_k \lambda_k +\frac{1}{\rho - 1}\sum_k \left(\min_j \left \{\frac{p_j+\lambda_k}{R_{k,j}}\right \}\right)^{1 - \rho} \label{dualobj}\\
\textrm{subject to}&\;\;\;  \pv, \lambdav \geq 0.
\end{align}
\end{subequations}
We next describe a convergent subgradient algorithm to approximate arbitrarily closely 
the solution to (\ref{target-dual}). We let $(\pv^{(i)}, \,\lambdav^{(i)})$ denote the value of the dual variables at the $i$-th subgradient iteration. 
The $(i+1)$-th iterate is given as the $i$-th iterate minus an appropriately scaled adjustment along the subgradient chosen 
at the current iteration \cite{boyd2003subgradient}. The subgradient algorithm comprises the following steps:
\begin{enumerate}
\item Initialize  $(\pv, \lambdav)$ to some arbitrary positive values $(\pv^{(0)}, \lambdav^{(0)})$ and let  $i = 0$.  
Also choose the number of iterations\footnote{Alternatively we could choose a stopping criterion for the algorithm.} $i_{\max}$ 
and the step sequence $s^{(i)} = \frac{a}{b+i}$ with appropriately chosen constants $a>0$,  $b\ge 0$.
\item Choose the subgradient for the $i$-th iteration, based on the objective function in (\ref{dualobj}) evaluated in the neighborhood of 
$(\pv^{(i)}, \lambdav^{(i)})$. In particular, let:\footnote{In case multiple $j$ indices maximize $R_{k,j}/(p^{(i)}_j+\lambda^{(i)}_k)$ 
any one of these can be used.}
\begin{subequations}
\label{sub-gradient-method}
\begin{equation}
j^{(i)}_k = \argmax_j 
\frac{R_{k,j}}{p^{(i)}_j+\lambda^{(i)}_k},
\end{equation}
and let
\begin{equation}
\Kc_j^{(i)}= 
\left\{ k\in \Kc;  \ \ \text{s.t.} \ \ j^{(i)}_k = j \right\}\ .  
\end{equation}
The $i$-th iteration subgradient is based on the derivative of the term
\begin{equation*}
\widecheck{G}^{(i)} (\pv, \lambdav) = \sum_j S_jp_j +\sum_k \lambda_k + F^{(i)}(\pv, \,\lambdav)
\end{equation*}
where
\begin{align*}
F^{(i)}(\pv, \,\lambdav)&=\frac{1}{\rho-1} 
\sum_k  \left(\frac{p_{j^{(i)}_k}+\lambda_k}{R_{k,j^{(i)}_k}}\right)^{1 - \rho}
=\frac{1}{\rho-1} 
\sum_j  \sum_{k\in\Kc_j^{(i)}} \left(\frac{p_j+\lambda_k}{R_{k,j}}\right)^{1 - \rho}.
\end{align*}
\item Taking derivatives of $\widecheck{G}^{(i)} (\pv, \lambdav)$ with respect to $p_j$ and $\lambda_k$, respectively, 
and the non-negativity constraint of $p_j$ and $\lambda_k$,  
the  corresponding subgradient iteration is given by 
\begin{align}
& p_j^{(i+1)} = \left [ p_j^{(i)} + s^{(i)} \left (\sum_{k \in \Kc^{(i)}_j}
\frac{R_{k,j}^{\rho-1}}{ \left(p^{(i)}_j+\lambda^{(i)}_k\right)^\rho}
-S_j \right ) \right ]^+ \label{APprice-update} \\
& \lambda_k^{(i+1)} = \left [ \lambda_k^{(i)} + s^{(i)} \left (
\frac{R_{k,j^{(i)}_k}^{\rho-1}}{ \left(p^{(i)}_{j^{(i)}_k}+\lambda^{(i)}_k\right)^\rho} -1 \right ) \right ]^+ \label{userprice-update}
\end{align}
\end{subequations}
\item If $i < i_{\max}$  increment $i$ by 1 and go to step 2, else stop.
\end{enumerate}
It can easily be verified that the formulas (\ref{sub-gradient-method}) also provide the corresponding subgradient 
algorithm iteration updates in the case $\gamma=1$. 

In the following, we let $(\pv^*, \lambdav^*)$ denote the dual variable values after a sufficiently large number of iterations of (\ref{sub-gradient-method})--({\ref{userprice-update}). Once $(\pv^*, \lambdav^*)$ have been obtained, we need to solve for the corresponding primal variables $\alphav^*$ in 
(\ref{OptfixedS}), thus obtaining the optimal association configuration and the corresponding optimal user throughputs. 
First, we first discuss the KKT conditions of (\ref{OptfixedS}). 
Then, in Section~\ref{sec:subgradient-primal}, we consider the primal variables solution.

\subsection{KKT conditions}  \label{KKT}

The convex program (\ref{OptfixedS}) is given in canonical form with linear inequality constraints
(\ref{equalityconst})--(\ref{ineq}). Therefore the Slater condition reduces to feasibility. This implies that strong duality holds and the KKT conditions including
the feasibility and the complementary slackness conditions are both necessary and sufficient for optimality. 
Noticing that all variables are non-negative (for a classical argument, see \cite[Th. 4.4.1]{gallager1968information}), by taking the partial derivatives of $L(\alphav, \rv,\betav,\pv,\lambdav)$ with respect to $r_k$ and $\alpha_{k,j}$ we obtain necessary and sufficient conditions 
for optimality in the form
\begin{eqnarray} 
\frac{\partial L}{\partial r_k}  =  \phi'_\gamma(r_k) - \beta_k & \leq & 0 \label{KKT1} \\
 \frac{\partial L}{\partial \alpha_{k,j}}  =  \beta_k R_{k,j} - p_j - \lambda_k & \leq & 0 \label{KKT2}
\end{eqnarray}
where inequalities (\ref{KKT1})-(\ref{KKT2}) must hold with strict equality for the strictly positive components $r_k$, $\alpha_{k,j}$, respectively, 
at the optimal points. The complementary slackness conditions are equivalently expressed as follows: at the optimal point, the inequalities 
\begin{eqnarray}
\sum_{k \in \Kc}\alpha_{k,j} - S_j & \leq & 0  \label{constJ}\\
 \sum_{j \in \Jc}\alpha_{k,j} - 1 & \leq  & 0, \label{constK} 
\end{eqnarray}
must hold with strict equality for all strictly positive components of $\pv$ and $\lambdav$, respectively. On the other hand, 
for the components $p_j = 0$ (resp., $\lambda_k = 0$), (\ref{constJ}) (resp., (\ref{constK})) may hold with strict inequality. 

Let  $(\alphav^o, \rv^o,\betav^o,\pv^o,\lambdav^o)$ denote an optimal point, i.e., a set of values 
for $(\alphav, \rv,\betav,\pv,\lambdav)$ that achieves the min (w.r.t. the dual variables) of the max (w.r.t. the primal variables) 
of (\ref{lagrangian}), and recall that we assumed $\gamma \geq 1$, implying that $r^o_k = 0$ for some $k$ cannot be an optimal 
point (see Remark \ref{remark-gamma1}). This implies that (\ref{KKT1}) must hold with equality at $(\alphav, \rv,\betav,\pv,\lambdav)=(\alphav^o, \rv^o,\betav^o,\pv^o,\lambdav^o)$.  Using the expression $\phi'_\gamma(x) = x^{-\gamma}$, the definition $r_k = \sum_{j \in \Jc_k} \alpha_{k,j} R_{k,j}$ and substituting $\beta_k = \phi'_\gamma(r_k)$ in (\ref{KKT2}) , (\ref{KKT1})-(\ref{KKT2}) reduce to
\begin{eqnarray} 
\sum_{j' \in \Jc_k} \alpha^o_{k,j'} R_{k,j'} & = &  \left ( \frac{R_{k,j}}{p^o_j + \lambda^o_k} \right )^{\rho} \;\;\; \forall \;\; j \in \Jc^*_k(\pv^o,\lambdav^o) \label{KKT41} \\
\sum_{j' \in \Jc_k} \alpha^o_{k,j'} R_{k,j'} & > &  \left ( \frac{R_{k,j}}{p^o_j + \lambda^o_k} \right )^{\rho} \;\;\; \forall \;\; j \notin \Jc^*_k(\pv^o,\lambdav^o), 
\label{KKT4}
\end{eqnarray}
where  the set $\Jc^*_k(\pv^o,\lambdav^o)$ is given by
\begin{equation} 
 \Jc^*_k(\pv^o,\lambdav^o) = \left\{j  \ : \  \frac{R_{k,j}}{p_j^o + \lambda_k^o}=   \max_{j'\in \Jc_k}\frac{R_{k,j'}}{p_{j'}^o + \lambda_k^o} \right\}.
 \label{max-bang-set}
\end{equation}
Summarizing, the consistency conditions for the optimality of the activity fractions at an optimal set of prices are as follows:
for each user $k \in \Kc$ and the sets of BSs defined in (\ref{max-bang-set}), we have
\begin{equation}
\left \{ \begin{array}{ll} 
\alpha^o_{k,j} > 0 & \textrm{for some} \;\; j \in \Jc^*_k(\pv^o,\lambdav^o) \\
\alpha^o_{k,j} = 0 & \;\;\; \forall \;\; j \notin \Jc^*_k(\pv^o,\lambdav^o). \end{array} \right .
\label{KKT5}
\end{equation}
Interpreting the quantity $\left(\frac{R_{k,j}}{p^0_j+\lambda^0_k}\right)^{\rho}$ as the ``{\it bang-per-buck}'' 
offered by BS $j$ to user $k$ (a term from economics \cite{devanur2008market}) 
at the {\it prices} $p^o_j$ and $\lambda^o_k$,  $\Jc^*_k(\pv^o,\lambdav^o)$ is the set of BSs offering the maximum bang-per-buck to user $k$.
Thus, the conditions (\ref{KKT41}) and (\ref{KKT5}) imply that,  at the optimum {\it prices} $\pv^o$ and $\lambdav^o$,
\begin{itemize}
\item the throughput of every user $k$ should be equal to the maximum bang-per-buck value offered by some BS in its neighborhood; 
\item every user $k$ can have a strictly positive activity fraction to only those BSs which offer the maximum bang-per-buck value, 
among the BSs in $\Jc_k$ (from which user $k$ is allowed to get service).
\end{itemize}

\subsection{Solving for  the Primal Variables}\label{sec:subgradient-primal}

We now describe how to use the solution $(\pv^*, \lambdav^*)$ of the dual problem (\ref{target-dual}) to solve for the primal variables. First we note that, using (\ref{mamax}) and
(\ref{mamamax}) we have 
\begin{equation}
\label{rk-star}
r^{*}_k =  \left (\max_j \left \{\frac{R_{k,j}}{p^{*}_j+\lambda^{*}_k}\right \} \right )^{\rho},\ \ \ \ \forall k\in \Kc\ .
\end{equation}
Hence, the optimal user throughputs are given by (\ref{rk-star}).
In order to calculate the optimal association configuration $\alphav^*$
from $(\pv^*, \lambdav^*)$, we can solve the KKT conditions.  In particular, we can choose any  feasible association configuration $\alphav^*$
satisfying
\begin{subequations}\label{KKT-useful}
\begin{eqnarray}
\alpha_{k,j}^* & = & 0  \;\;\; \forall \;\; j \notin \Jc^*_k(\pv^*,\lambdav^*)  \\
\sum_{j' \in \Jc^*_k(\pv^*,\lambdav^*)} \alpha^*_{k,j'} R_{k,j'} & > &  \left ( \frac{R_{k,j}}{p^*_j + \lambda^*_k} \right )^{\rho} \;\;\; \forall \;\; j \notin \Jc^*_k(\pv^*,\lambdav^*)  \\
\sum_{j' \in \Jc^*_k(\pv^*,\lambdav^*)} \alpha^*_{k,j'} R_{k,j'} & = &  \left ( \frac{R_{k,j}}{p^*_j + \lambda^*_k} \right )^{\rho} \;\;\; \forall \;\; j \in \Jc^*_k(\pv^*,\lambdav^*)  \\
p_j^* \big(\sum_{k \in \Kc}\alpha_{k,j}^* - S_j\big) & = & 0 \;\;\; \forall \;\; j \in \Jc \\
\lambda_k^* \big(\sum_{j \in \Jc} \alpha_{k,j}^* - 1\big) & = & 0  \;\;\; \forall \;\; k \in \Kc. 
\end{eqnarray} 
\end{subequations}

However in practice, the dual subgradient algorithm yields dual variables that differ from their optimal value by some very small numerical error, due to finite 
machine precision and finite number of iterations. Hence, the system of KKT conditions above may not have a solution when $\pv^*$ and $\lambdav^*$ are numerically calculated. We therefore propose a numerically stable approach that always yields a feasible association configuration and, 
in particular,  yields the exact optimal $\alphav^*$ when $\pv^*$ and $\lambdav^*$ are exactly at their optimal point (see Lemma \ref{lemlem1} below). 

As noticed before, the optimal throughput values $\rv^*$ are given by (\ref{rk-star}).
Define the ratios $\widetilde{R}_{k,j} = R_{k,j} / r^*_k$, and consider
the quantities $f_k(\alphav) = \sum_{j \in \Jc_k} \alpha_{k,j} \widetilde{R}_{k,j}$, for $k \in \Kc$. 
By construction, there exists an optimal feasible association configuration $\alphav^*$
such that $f_k(\alphav^*) = 1$, i.e., there is an optimal point where the (linear) functions $f_k(\cdot)$ are equal to $1$, 
for all $k \in \Kc$.  This suggests that $\alphav^*$ can be found as the solution of the LP:
\begin{subequations}\label{Optmaxmin}
\begin{align}
 \textrm{maximize} & \;\;\; \theta \label{maxminutil}\\
 \textrm{subject to} & \;\;\; \theta \leq f_k(\alphav), \; \forall \; k \in \Kc \label{maxmineqc} \\
&  \;\;\; \sum_{k \in \Kc}\alpha_{k,j} \leq S_j, ~\forall~j \in \Jc \label{maxmincJ}\\
& \;\;\; \sum_{j \in \Jc}\alpha_{k,j} \leq 1,~\forall~k \in \Kc \label{maxmincK} \\
& \;\;\; \alpha_{k,j} \geq 0,~\forall~k \in \Kc,~j \in \Jc. \label{maxminineq}
\end{align} 
\end{subequations}
We have:

\begin{lemma} \label{lemlem1}
If the throughputs $r_k^*$ given by (\ref{rk-star}) correspond to the exact optimal solution of the NUM problem (\ref{OptfixedS}), then
the solution of (\ref{Optmaxmin}) is the corresponding optimal feasible association configuration.
\end{lemma}

\begin{IEEEproof}
Notice that (\ref{Optmaxmin}) maximizes the minimum $f_k(\cdot)$ by maximizing a common lower bound $\theta$ subject to the feasibility of the association configuration. 
Let $\widehat{\alphav}$ denote the solution of $(\ref{Optmaxmin})$ and let $\theta_{\max} = \min_k f_k(\widehat{\alphav})$ denote the achieved maximum value of 
the common lower bound. 
Since, by construction, there exist a feasible configuration $\alphav^*$ for which $f_k(\alphav^*) = 1$ for all $k \in \Kc$, 
there are two possible cases: 1) $\theta_{\max} < 1$, or 2) $\theta_{\max} \geq 1$.
Case 1) is impossible, otherwise $\widehat{\alphav}$ could be improved to $\alphav^*$, contradicting the assumption that $\widehat{\alphav}$ is the solution of  (\ref{Optmaxmin}).
Case 2) can only hold with equality. In fact, if it held with strict inequality, we would have $f_k(\widehat{\alphav}) > 1$ for all $k$, implying 
\[ \sum_{j \in \Jc_k} \widehat{\alpha}_k R_{k,j} > \sum_{j \in \Jc_k} \alpha^*_k R_{k,j} = r^*_k, \;\; \forall \; k \in \Kc. \]
Since the network utility function $U(\cdot)$ is componentwise increasing, this means that there exists a feasible association configuration $\widehat{\alphav}$ yielding better utility than the optimal $\alphav^*$, thus leading to a contradiction. 
It follows that it must be $\theta_{\max} = 1 = f_k(\widehat{\alphav})$ for all $k \in \Kc$, implying 
$\widehat{\alphav} = \alphav^*$ since, by the system model setup,  $R_{k,j} > 0$ for all $j \in \Jc_k$. 
\end{IEEEproof} 

\begin{remark}
In order to derive a centralized association/scheduling policy that yields the activity fractions that are the solution of (\ref{OptfixedS}), 
some association and activation sequences must be found with empirical activity fractions converging (in the limit for $T \rightarrow \infty$) 
to $\alphav^*$.  Theorem~\ref{th-feasibility} guarantees that this is possible.
However, finding such sequences is not easy in general. From the proof of Theorem 
\ref{th-feasibility} in Appendix \ref{proof:feasibility}, 
we can see that this is equivalent to finding integer scheduling configurations such that $\alphav^*$ can be written as their
convex combination. Unfortunately, this is again a combinatorial problem that may be hard to solve in general.
\hfill $\lozenge$
\end{remark}

Motivated by this consideration, in the next section we study a class of schemes that yield 
{\em unique association configurations} and fully decentralized user-centric association/scheduling policies. 
Yet, somehow surprisingly, these schemes are shown to perform close to the globally optimal centralized solution.

\section{Distributed User-Cell Association Algorithms}\label{sec:user centric}

In this section, we focus on user-cell association algorithms where each user makes its own association 
decisions in a selfish way, i.e., based on its own user-centric utility function. In particular, we consider a class of such schemes
where the user-centric utility function is the user throughput, $r_k$, and each BS applies a local version of the NUM (\ref{OptfixedS})
in order to allocate its transmission resources (time-frequency slots) among its associated users.

Letting $\Kc_j$ denote the set of users uniquely associated to BS $j$,  the service policy at each BS $j \in \Jc$ 
solves the following local NUM problem:
\begin{subequations}\label{NUM-local}
\begin{align}
 \textrm{maximize} & \;\;\; \sum_{k \in \Kc_j} \phi_\gamma( \alpha_{k,j} R_{k,j})   \label{maxutil-local}\\
 \textrm{subject to} &  \;\;\; \sum_{k \in \Kc_j} \alpha_{k,j} \leq S_j,  \label{matchingconstJ-local}\\
& \;\;\; 0 \leq \alpha_{k,j} \leq 1, ~\forall~k \in \Kc_j. \label{ineq-local}
\end{align} 
\end{subequations}
The solution of this problem is given by:
\begin{theorem} \label{BS-NUM}
For $\gamma \geq 1$ (i.e, $\rho \leq 1$), without loss of generality assume that\footnote{If this is not the case, 
the users must be sorted in non-increasing order with respect to the values $R_{k,j}^{\rho-1}$ by some permutation $\pi$, and the statement of the theorem is
valid by replacing the user index $k$ with its permuted version $\pi(k)$.} 
\begin{equation} \label{sort-users}
R_{1,j}^{\rho-1} \geq R_{2,j}^{\rho-1} \geq \ldots \geq R_{|\Kc_j|,j}^{\rho-1}.
\end{equation}
Setting $R_{0,j}^{\rho-1} = \infty$ and $R_{|\Kc_j|+1,j}^{\rho-1} = 0$, let $k^* \in \{1, \ldots, |\Kc_j|\}$ be such that 
\begin{equation} \label{KKT-local}
R_{k^*-1,j}^{\rho-1} \geq \frac{\sum \limits_{k=k^*}^{|\Kc_j|}R_{k,j}^{\rho-1}}{S_j-k^*+1}  > R_{k^*,j}^{\rho-1}.
\end{equation} 
Then, the solution of (\ref{NUM-local}) is given by
\begin{equation}\label{alpha-final}
\alpha_{k,j} = \left \{ \begin{array}{ll}
1, & \;\;\; \mbox{for} \;\; 1\leq k \leq k^*-1 \\
\frac{(S_j-k^*+1)R_{k,j}^{\rho-1}}{\sum \limits_{k=k^*}^{|\Kc_j|}R_{k,j}^{\rho-1}} & \;\;\; \mbox{for} \;\; k^* \leq k \leq |\Kc_j| \end{array} \right .
\end{equation}
\end{theorem}

\begin{IEEEproof} 
The proof is given in Appendix \ref{proof-BS-NUM}. Here, we just note that an index $k^*$ satisfying (\ref{KKT-local}) always exists.
When this is not unique (e.g., in the case $|\Kc_j| = S_j$ and $\rho = 1$), any choice of $k^*$ satisfying (\ref{KKT-local}) yields the same
optimal value of the objective function.
\end{IEEEproof}

\begin{remark} 
For the particularly important case of PF ($\gamma=1$) the solution of Theorem \ref{BS-NUM} reduces to: 
\begin{equation}\label{alpha-final-pfs}
\alpha_{k,j} = \left \{ \begin{array}{ll}
\frac{S_j}{|\Kc_j|}, & \;\;\; \forall \;k \in \Kc_j \;\; \mbox{if} \;\; S_j \leq |\Kc_j| \\
1 & \;\;\; \forall \;k \in \Kc_j  \;\; \mbox{if} \;\; S_j > |\Kc_j| \end{array} \right .
\end{equation}
\hfill $\lozenge$
\end{remark}

\begin{remark}
Under the condition
\begin{equation}\label{heavy-load-new}
\frac{S_j R_{k,j}^{\rho-1}}{\sum_{k'\in \Kc_j} R_{k',j}^{\rho-1}}  \leq 1, \;\;\; \forall \;\; j \in \Jc \;\; \mbox{and} \;\; k \in \Kc_j, 
\end{equation}
the solution of Theorem \ref{BS-NUM} simplifies to:
\begin{equation}  \label{alpha-local}
\alpha_{k,j} = \frac{S_{j} R_{k,j}^{\rho - 1}}{\sum_{k' \in \Kc_j} R_{k',j}^{\rho - 1}}, \;\;\; \forall \;\; k \in \Kc_j.
\end{equation} 
The term $\sum_{k' \in \Kc_{j}} R_{k',j}^{\rho - 1}$ in the denominator of (\ref{alpha-local}) can be interpreted as a measure of the {\it load} 
of  BS $j$.  For example, for the case of proportional fairness this quantity is simply equal to $|\Kc_{j}|$, i.e., 
the number of users  uniquely associated with BS $j$. Condition (\ref{heavy-load-new}) shall be referred to as the {\em heavy load} condition. 
Since we are interested in the performance of the network in heavy load conditions,\footnote{If the network is lightly loaded, 
some spatial dimensions at some BSs may be under-utilized, meaning that not all the downlink streams need to be used at each slot time.
In this case, the network has more downlink capacity than needed, and other problems beyond NUM become relevant, as for example
the transmit power minimization subject to given target per-user throughputs (see for example \cite{athanasiou2013optimizing}). 
This non-heavy-loaded regime is not the focus of this paper, and its study in the context of massive MIMO
is left for future work.}  in the following we shall assume that (\ref{heavy-load-new}) holds  for all BSs $j \in \Jc$. \hfill $\lozenge$
\end{remark}

Since the activity fractions $\{\alpha_{k,j}\}$ are fixed by the local NUM policy (\ref{alpha-local}), we shall not distinguish any longer between 
the partition and the induced unique association configuration. In particular, we have:

\begin{definition} \label{def:valid-partition}
{\bf Valid partition:}
A partition $\{\Kc_j : j \in \Jc\}$ of the user set $\Kc$ is {\em valid} if it corresponds to a feasible unique association 
configuration. In particular, the corresponding association vector $\jv$ has components $j_k \in \Jc_k$ defined by 
$j_k = j \Leftrightarrow k \in \Kc_j$, for all $k \in \Kc$. \hfill $\lozenge$
\end{definition} 

For a given valid partition $\{\Kc_j\}$, the throughput of any user $k$ is given by  $r_k = \alpha_{k,j_k} R_{k,j_k}$.

\subsection{User-centric association games}

Following \cite{chiang2013ratselection}, the user-centric association algorithms considered in this work can be studied 
in the framework of non-cooperative association games. In particular, the normal-form association game is defined by:
\begin{itemize}
\item Players: the users $k \in \Kc$.
\item Action space: each user $k$ has an action set $\Jc_k$ where action $j_k \in \Jc_k$ corresponds to the decision of
user $k$ to associate uniquely to BS $j_k$. Therefore, the joint action set of all users is the Cartesian product
$\Ac = \Jc_1 \times \cdots \times \Jc_K$.
\item Payoff functions: the payoff function of user $k$ is its throughput $r_k =  \alpha_{k,j_k} R_{k,j_k}$, where $R_{k,j_k}$ is a fixed value that depends on 
the massive MIMO downlink scheme employed (see Appendix \ref{massiveMIMO}), and $\alpha_{k,j_k}$ is given by (\ref{alpha-local}). 
\end{itemize}
Since according to (\ref{alpha-local}) $\alpha_{k,j}$ is a function of $\Kc_{j}$,
it follows that the user throughputs $r_k$ are functions of the joint action $\jv \in \Ac$, i.e., 
the payoff functions are maps $r_k : \Ac \rightarrow \RR_+$. In order to stress this dependency, 
we shall use the notation $r_k(\jv)$. 

A unique association configuration $\jv = (j_1, \ldots, j_K)$ is said to be a {\it pure Nash equilibrium}  if, 
for all $k \in \Kc$,  we have $r_k\left(j_1, \ldots, j_k,\ldots, j_K\right) \geq r_k\left(j_1, \ldots, j, \ldots, j_K\right)$ for any 
$j \in \Jc_k$.  In other words, no user has an incentive to change unilaterally its association while all other users stay unchanged.

Before discussing a specific user centric algorithm in terms of its on-line decentralized implementation, 
let's examine the global optimality properties of the Nash equilibria of the related association game defined above.  
Suppose that there exists a valid partition $\{\Kc_j : j\in \Jc\}$ for which (\ref{heavy-load-new}) holds and such that:
\begin{align}
& \frac{S_{j_k} R_{k,j_k}^{\rho}}{\sum_{k'\in \Kc_{j_k}} R_{k',j_k}^{\rho-1}} > \frac{S_\ell R_{k,\ell}^{\rho}}{\sum_{k'\in \Kc_\ell} R_{k',\ell}^{\rho-1}}, \;\;\; 
\forall \;\; k \in \Kc, \;\; \mbox{and} \;\; \ell \in \Jc_k \;\; \mbox{with} \;\; \ell \neq j_k. \label{max-bang-cond}
\end{align}
Then, setting the dual variables of the global NUM problem as
\begin{equation}  \label{unique-prices}
p_j^* = \left ( \frac{1}{S_j} \sum_{k' \in \Kc_j} R_{k',j}^{\rho-1} \right )^{1/\rho}, \;\;\; \mbox{and} \;\; \lambda_k^* = 0,
\end{equation}
we can easily verify that the KKT conditions (\ref{KKT-useful}) are satisfied with
\[ \alpha_{k,j}^* = \left \{ \begin{array}{ll}
\frac{S_j R_{k,j}^{\rho-1}}{\sum_{k'\in \Kc_j} R_{k',j}^{\rho-1}} & \;\;\; \mbox{for} \;\; j = j_k \\
0 & \;\;\; \mbox{for} \;\; j \neq j_k, \end{array} \right . \]
which coincide with (\ref{alpha-local}) and, using  (\ref{rk-star}), yield the user throughputs $r_k^* = \left ( \frac{R_{k,j_k}}{p_{j_k}^*} \right )^\rho = \frac{S_{j_k} R_{k,j_k}^{\rho}}{\sum_{k'\in \Kc_{j_k}} R_{k',j_k}^{\rho-1}}$. Obviously, these throughputs are the same obtained by applying the local NUM policy (solution of (\ref{NUM-local})) 
at each BS $j$,  under the unique association given by $\{\Kc_j : j\in \Jc\}$.  Since the KKT conditions are necessary and sufficient, we conclude that the valid partition
$\{\Kc_j : j\in \Jc\}$ combined with the local NUM policy (\ref{alpha-local}) yields a globally optimal unique association 
configuration for the network-wide NUM problem (\ref{OptfixedS}).  Also, notice that the inequalities (\ref{max-bang-cond}) imply the pure-strategy Nash equilibrium
\begin{equation} \label{nash}
\frac{S_{j_k} R_{k,j_k}^{\rho}}{\sum_{k'\in \Kc_{j_k}} R_{k',j_k}^{\rho-1}} \geq 
\frac{S_\ell R_{k,\ell}^{\rho}}{R_{k,\ell}^{\rho-1} + \sum_{k'\in \Kc_\ell} R_{k',\ell}^{\rho-1}}, \;\;\;
\forall \;\; k \in \Kc, \;\; \mbox{and} \;\; \ell \in \Jc_k \;\; \mbox{with} \;\; \ell \neq j_k.
\end{equation}
Summarizing, we have proved:

\begin{lemma} \label{lemma-stupid}
If a valid partition $\{\Kc_j : j \in \Jc\}$ satisfies  (\ref{heavy-load-new}) and (\ref{max-bang-cond}), then the corresponding 
association $\jv$  is a pure-strategy Nash equilibrium of the decentralized association 
game. Furthermore, such Nash equilibrium corresponds to the global optimum of the network-wide 
NUM problem (\ref{OptfixedS}). \hfill $\blacksquare$
\end{lemma}



Now, suppose that the association game has a pure-strategy Nash equilibrium. 
Hence, there exists a unique association $\jv$ such that (\ref{nash}) holds. 
Arguing in the reverse direction of the argument leading to Lemma \ref{lemma-stupid}, we observe that if the 
heavy-loaded condition (\ref{heavy-load-new}) holds, then the KKT conditions (\ref{KKT-useful}) of the global problem are ``almost'' satisfied. 
For example, in the case PF ($\gamma = 1$), the Nash equilibrium conditions are given by 
\begin{equation} \label{nash-PF}
\frac{S_{j_k} R_{k,j_k}}{|\Kc_{j_k}|} \geq 
\frac{S_\ell R_{k,\ell}}{|\Kc_\ell| + 1}, \;\;\;
\forall \;\; k \in \Kc, \;\; \mbox{and} \;\; \ell \in \Jc_k \;\; \mbox{with} \;\; \ell \neq j_k.
\end{equation}
When $|\Kc_j| > S_j \gg 1$ for all $j \in \Jc$ (heavy-loaded system, typical in the case of massive MIMO networks), 
we have that the ``+1'' in the denominator of the promised rate terms is negligible with respect to the set size $|\Kc_\ell|$, 
and therefore the Nash equilibrium condition and the KKT conditions (\ref{KKT-useful}) {\em almost} coincide. 
This means that, for heavy-loaded systems, a decentralized user-centric system operating 
at its Nash equilibrium is also very close to the global optimum of the network-wide NUM problem. 

\subsection{User-centric decentralized on-line algorithms}

For a decentralized on-line implementation of the user-centric association scheme, several variants have been proposed. \footnote{For example, \cite{chiang2013ratselection} discusses also the case where two subsets of BSs operate according to different local NUM policies; one subset operates according to PF (equal air-time) and another subset operates according to HF (equal throughput).} 
Here, and in our simulations, we restrict to a very simple scheme that requires only local information at the users. 
In practice this can be easily obtained from the BSs through the ``beacon-stuffing'' approach~\cite{chandra2007beacon}, 
where each BS advertises the required information while broadcasting its beacon signal. 
In the proposed scheme, starting from a current association $\jv$, each user $k$ compares its current 
throughput $r_k(\jv)$ with the highest promised throughput 
\begin{equation}\label{promised}
\widehat{r}_k = \max_{\ell \in \Jc_k \setminus j_k} \; \alpha_{k,\ell} R_{k,\ell},
\end{equation}
where $\alpha_{k,\ell}$ is given by (\ref{alpha-final}).
If $\widehat{r}_k > r_k(\jv)$, then user $k$ changes its association to BS $\widehat{\ell}$ achieving the max in (\ref{promised}), 
with some fixed probability $\pi  \in (0,1)$. Otherwise, user $k$ keeps its current association. 

This algorithm evolves according to a discrete-time Markov Chain, with state space $\Ac$ (i.e., the joint action space of the 
related association game). 
Since $\Ac$  is a finite set and since every state has a self-transition of positive probability, the chain is aperiodic. 
Furthermore, it is immediate to see that the pure-strategy Nash equilibria of the game are absorbing states. 
If any state $\jv \in \Ac$ communicate with a each equilibrium (i.e.,  there is a path of positive probability from $\jv$ to a Nash equilibrium state), 
then the only persistent classes are the Nash equilibria and all other states are transients. 
In this case, we have that the algorithm converges to a Nash equilibrium with probability $1$. 
 
An {\it improvement path} in the association game consists of a sequence of unique association 
configurations $\jv$, each differing from the preceding one in a single component only, such that the change of strategy 
in that component increases the throughput of the corresponding user. If every improvement path is finite, then the game is said to have the 
{\em finite improvement path} property, which implies that any state communicates with a Nash equilibrium state 
(see \cite{chiang2013ratselection} and references therein). For $\gamma=1$ (PF) and $\gamma \rightarrow \infty$ (HF),  it is known that such property holds
\cite{chiang2013ratselection}, implying that the user-centric decentralized association algorithm 
converges with probability 1 to a pure-strategy Nash equilibrium. For what was said before, 
this Nash equilibrium is very close to the global NUM optimum, when the network is in the heavy-loaded condition (\ref{heavy-load-new}). 
This is confirmed by extensive simulations, some of which are reported in Section \ref{sec:simulation}.  
Furthermore, we always observed convergence also for other values of $\gamma \in (1, \infty)$.  
Therefore, we conjecture that pure-strategy Nash equilibria exist with high probability for random network topologies and 
arbitrary fairness factor $\gamma$,  although proving the finite improvement path property has been, so far, elusive.

\begin{remark} \label{time-varying-networks}
In practice, networks have a (slowly) time-varying topology due to user motion across the coverage area, and transitions due to users joining or leaving the system.  Hence, the association must be continuously updated. We notice that the user-centric randomized scheme proposed here, with positive 
switch probability $\pi$, is naturally suited to this purpose, allowing association switching as the user peak rates evolve in time due to user mobility or the BSs load changes due to users joining or leaving the system. Also, it is practical to introduce some hysteresis in order to prevent too frequent association switches
(which incur some protocol cost) and too wild fluctuations of the user peak rates. However, these  practical considerations go beyond the scope of 
this paper. \hfill $\lozenge$
\end{remark}

\section{ Numerical Examples and Concluding Remarks} \label{sec:simulation}

In this section we present a comparative evaluation of the user-centric load-balancing scheme considered in this paper, and a heuristic scheme based on maximum peak-rate association.
In particular, there is no ``standard'' commonly accepted way to perform user-BS association in a network employing multiuser MIMO. Here, as a term of comparison we have chosen a naive {\em maximum peak rate} association scheme, i.e., user $k$ associates  with BS $j(k) = \argmax_{j \in \Jc_k} R_{k,j}$.  In the massive MIMO regime, the user peak rates converge to the deterministic limit (\ref{user-rates-massive-MIMO}) which depends only on the individual user SINR terms (\ref{sinr-zfbf}) or (\ref{sinr-cbf}), which can be assumed to be known. 
Hence, the Max peak-rate association scheme can be easily implemented in the massive MIMO case. After the users associate with the BSs using the Max 
peak-rate decision, the BSs locally implement the $\gamma$-fairness policy according to (\ref{alpha-final}). We remark that the Max-RSRP scheme mentioned in Section \ref{sec:intro} does not apply since, in general, the SINR achieved by any user with multiuser MIMO downlink spatial multiplexing 
depends on the channel matrix realization and on the set of simultaneously scheduled users. 
 
\subsection{Experiment 1}
In this experiment, we consider a network topology formed by a 900m$\times$1800m rectangular region with several small-cell BSs and two macro BSs whose locations are fixed throughout all of the simulation runs. As shown in Fig.~\ref{layout-2macro}, the two macro BSs (indicated by the $\square$) are located in the centers of the two 900m$\times$900m square sub-regions comprising the 900m$\times$1800m rectangular area, and $40$ small cell BSs (indicated by $\circ$'s) are uniformly distributed in the region. The number of users (indicated by $*$'s) and their locations change across different simulation runs, and are generated according to a non-homogeneous Poisson point process with higher density in a central region around each of the Macro BSs, as shown in Fig.\ref{layout-2macro}. 

\begin{figure}[ht]
\centering
\includegraphics[scale = 0.5]{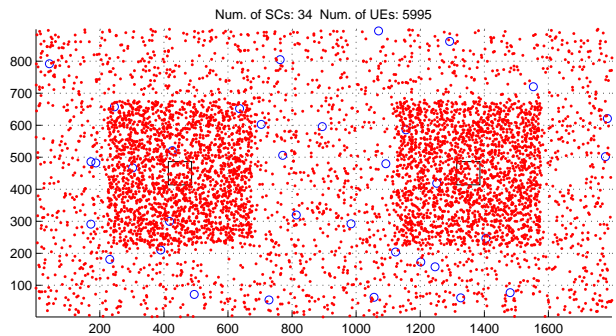}
\caption{Wireless network with 2 Macro BSs and several small cell BSs}
\label{layout-2macro}
\end{figure}

The macro BS has $M = 100$ antennas and serves user sets of size $S=10$, with $46$dBm transmission power. Each small-cell BS has $M=40$ antennas and serves user sets of size  $S=4$, with transmission power of $35$dBm. The pathloss from the macro BS to a user and from a small-cell BS to a user is given by $\frac{1}{1+(\frac{d}{40})^{3.5}}$ and by $\frac{1}{1+(\frac{d}{40})^{4}}$, respectively\footnote{A greater pathloss exponent ($4$) is used for small-cell BSs, in order to take into account the fact that the macro-BS antennas are at higher elevation.}, with $d$ representing the BS-user distance (assuming a torus wrap-around model to avoid boundary effects). 

We calculate the peak rates $R_{k,j}$ using the formulas (\ref{user-rates-massive-MIMO}) and (\ref{sinr-zfbf}) for ZFBF with pilot contamination. We assume that each Macro BS uses the same set  of $S=10$ pilots which are mutually orthogonal while the small cell BSs use a different set of $S=4$ pilots which are mutually orthogonal. 

In Figs. \ref{perctlgamma1}-\ref{statisticgamma1}, we compare the performance of the proposed centralized and distributed algorithms with the Max peak-rate association scheme. We choose a constant switching probability $\pi=0.1$ when simulating the distributed algorithm. For every realization of the layout similar to Fig. \ref{layout-2macro}, we calculate the {\it throughput statistics} 1) the $5\%$ percentile throughput, 2) the geometric mean of user 
throughputs and 3) the arithmetic mean of user throughputs and then plot the CDFs of these quantities over $100$ realizations for the case of $\gamma = 1$ (PF scheduling) in Figs. \ref{perctlgamma1}, \ref{gmeangamma1} and \ref{ameangamma1} respectively. We have run the centralized solution computed via the method in Section \ref{sec:subgradient} and, remarkably, the performance of the randomized distributed user-centric algorithm is almost indistinguishable from the performance of the (optimal) centralized solution, as we have argued to hold for highly-loaded systems in Section \ref{sec:user centric}. Furthermore, in Fig.~\ref{statisticgamma1}, we compare the performance of the distributed algorithm with the Max peak-rate scheme by calculating the ratio between the throughput statistic of the distributed algorithm and the throughput statistic of the Max peak-rate for every realization and then plot the CDF of the ratio over $100$ realizations. Fig.~\ref{statisticgamma1} reveals the fact that distributed algorithm results in superior performance in terms of the $5\%$ percentile throughput, and the geometric mean. For instance, half of the realizations observe more than $30 \%$ gain in $5$ percentile throughput with the distributed algorithm over the Max peak-rate scheme. Nevertheless, the Max peak-rate achieves higher average throughput, since the PF fairness function imposes to serve all users in a proportionally fair way across the network, while Max peak-rate does not. 

\begin{figure}[t]
\subfloat[]{
\includegraphics[scale=0.42]{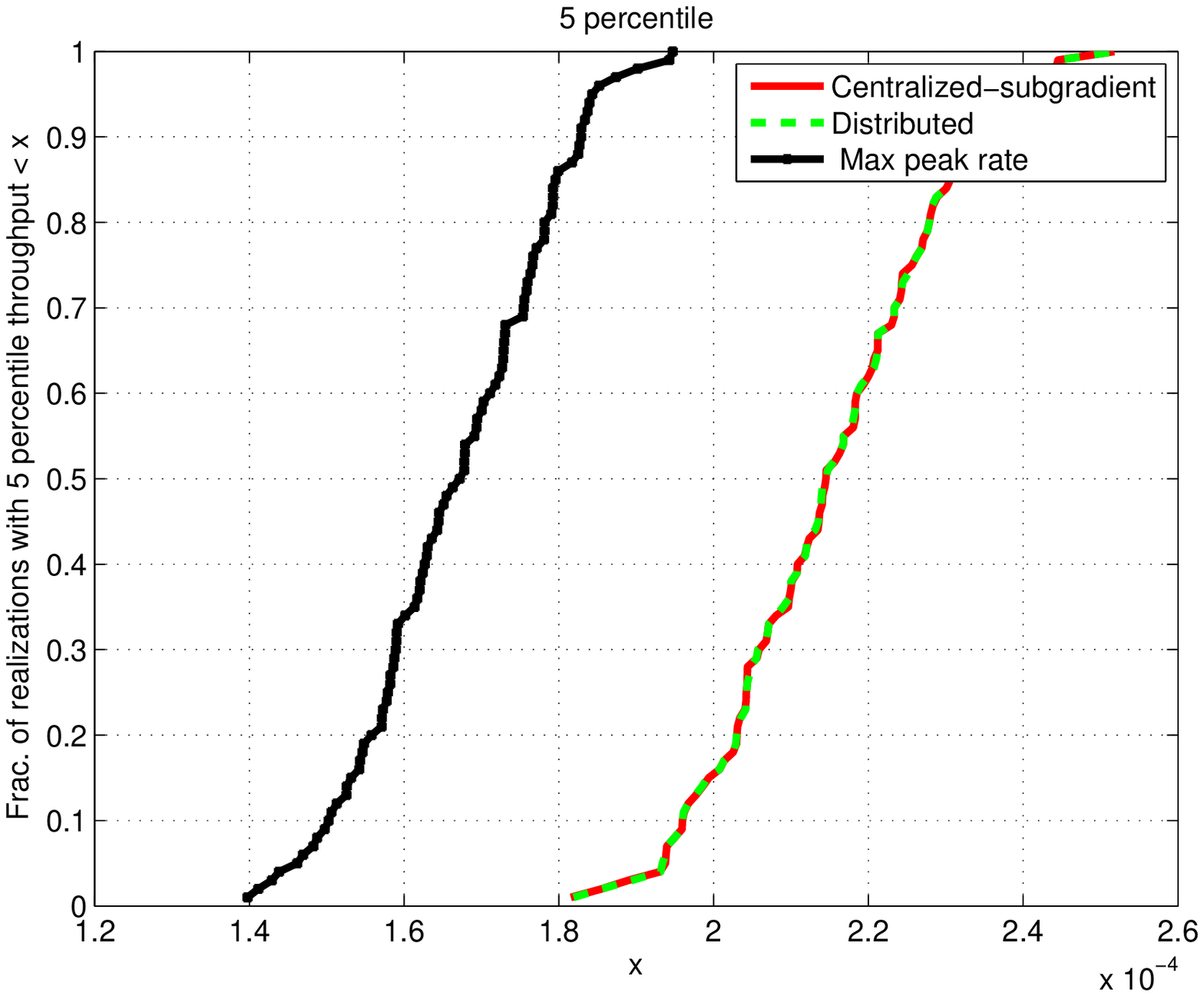}
\label{perctlgamma1}
}
\subfloat[]{
\includegraphics[scale=0.42]{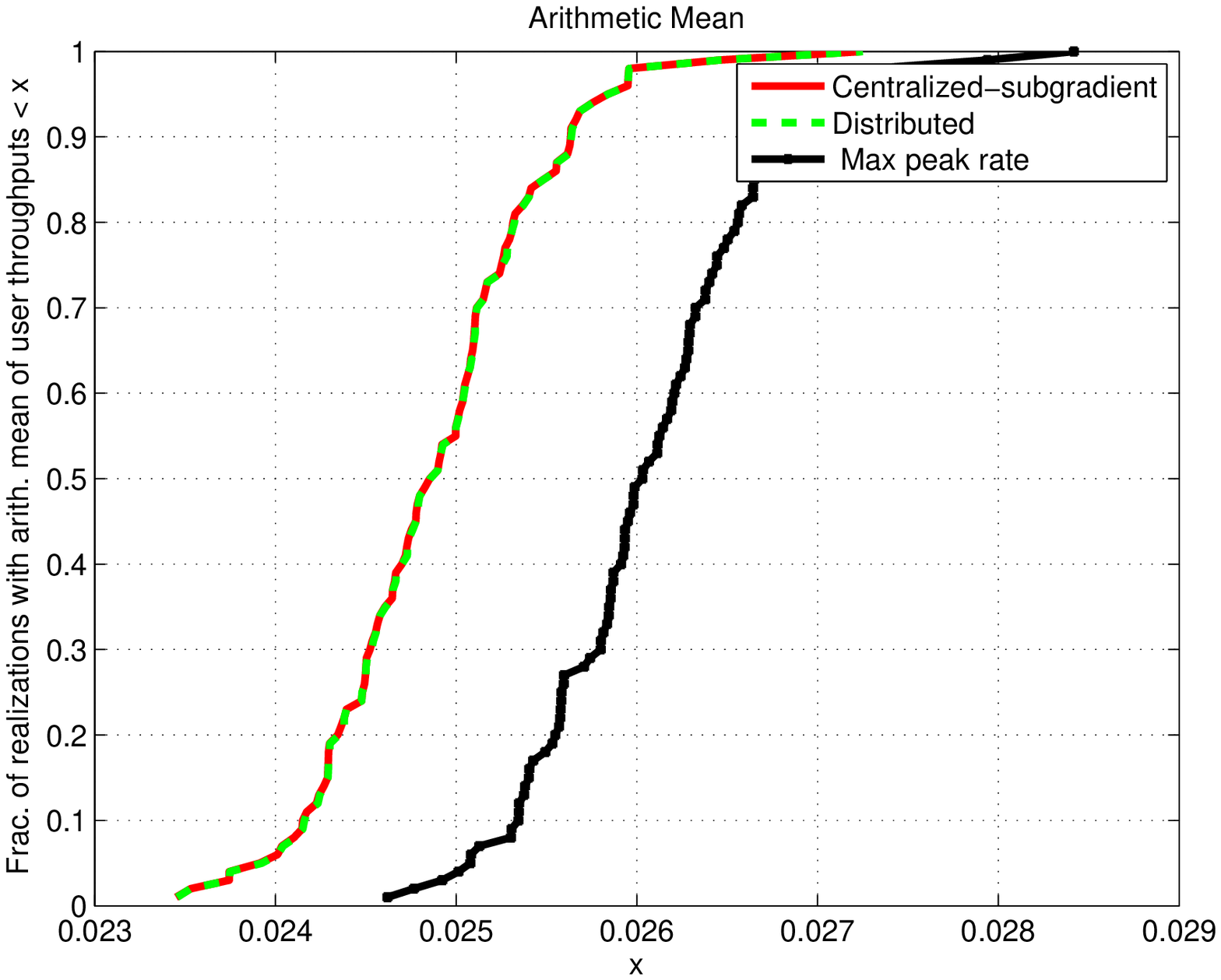}
\label{ameangamma1}
}\\
\centering
\subfloat[]{
\includegraphics[scale=0.42]{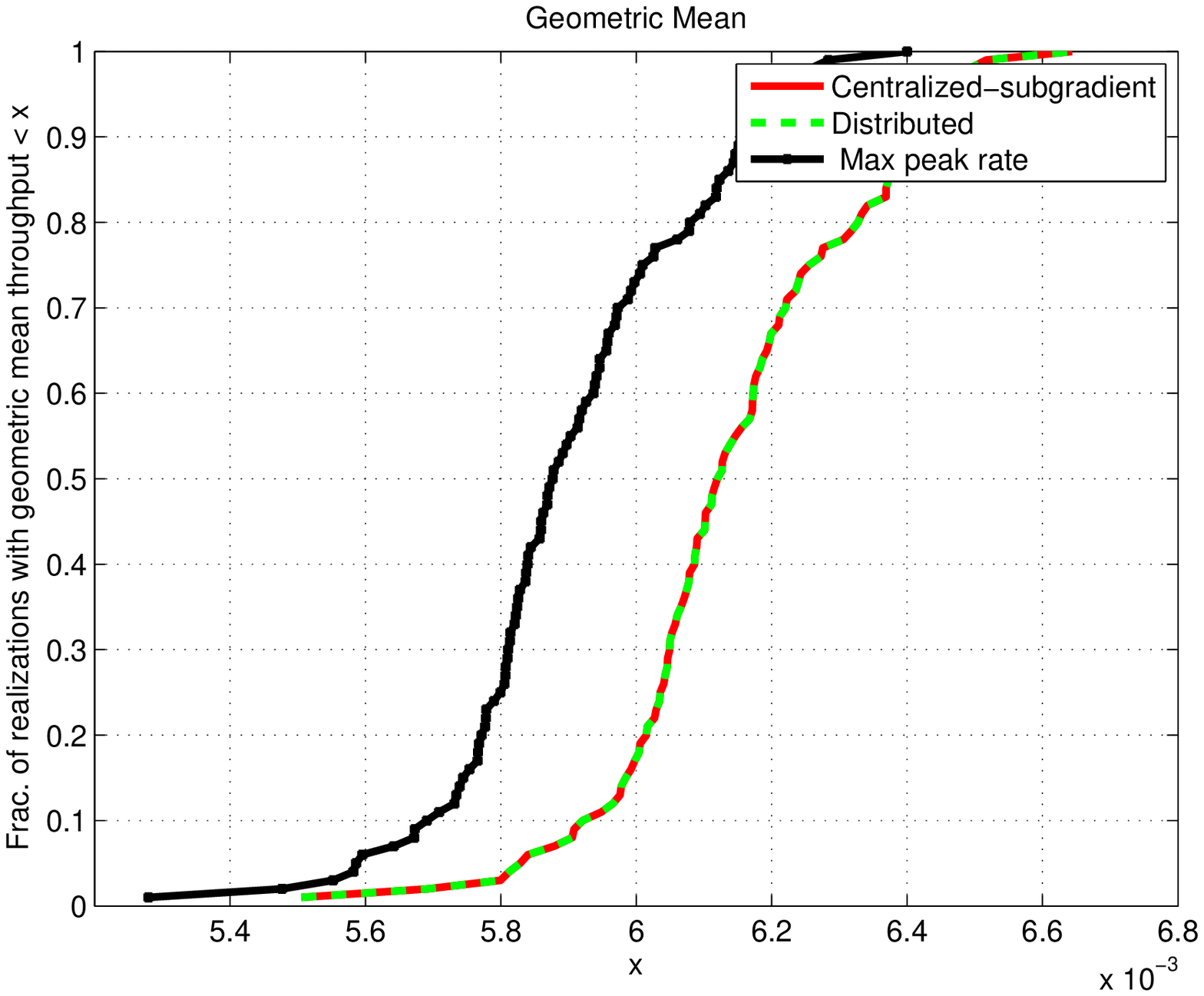}
\label{gmeangamma1}
}
\subfloat[]{
\includegraphics[scale=0.42]{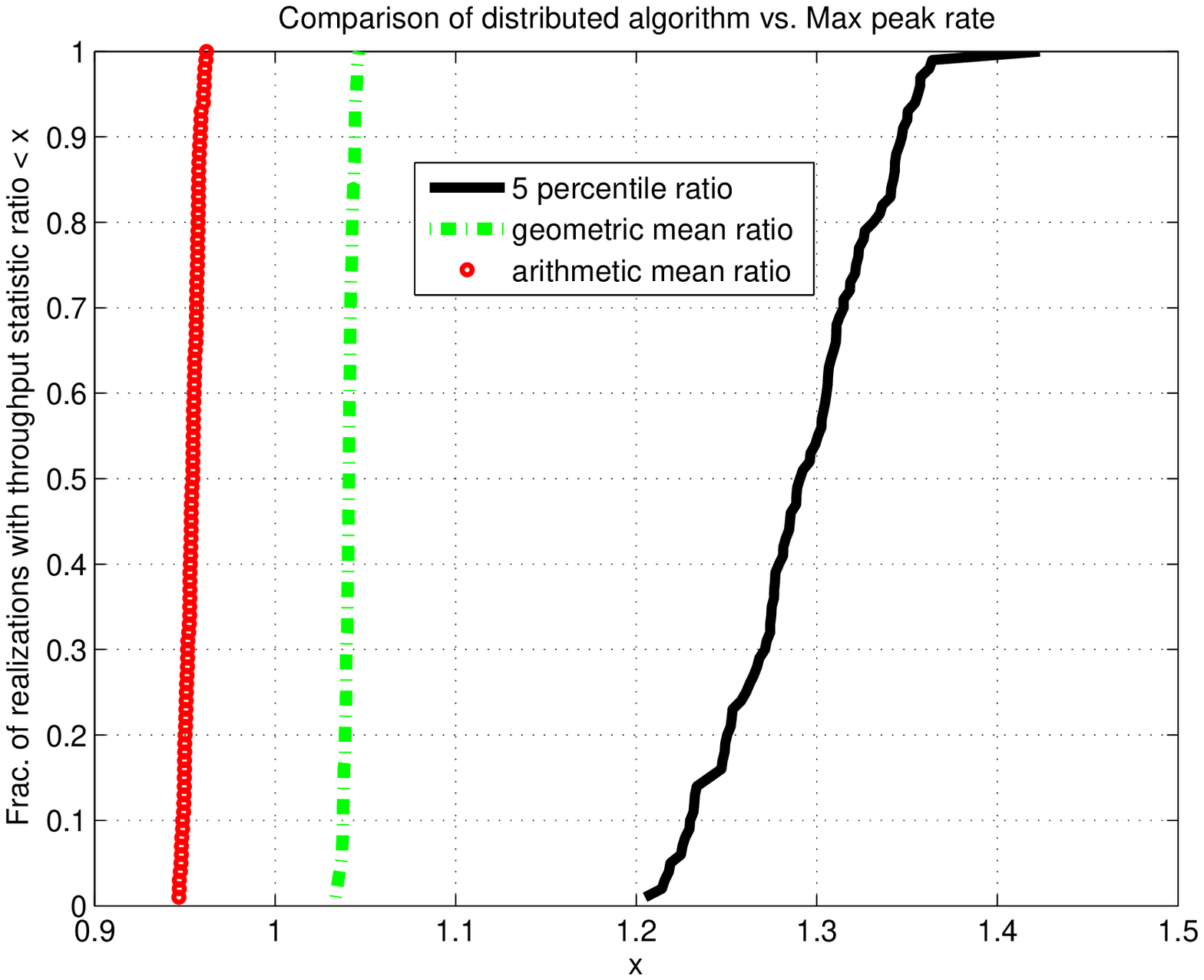}
\label{statisticgamma1}
}
\caption{Performance comparison of various algorithms for $\gamma=1$ and the layout of Fig.~\ref{layout-2macro}.}
\end{figure}

\subsection*{Experiment 2: 3GPP HetNet Model}

In this experiment, we conduct simulations in a more realistic network topology as shown in Fig.~\ref{layout} which is compliant with the layout specified for small cell heterogenous networks in 3GPP standardization \cite{3GPP}. In particular, we have a cellular layout with $7$ Macro cells (indicated by the $\square$) with each macro cell consisting of $3$ {\it hot zones}. A {\it hot zone} is a geographical area where the concentration of users (indicated by the green $+$'s) is much higher than the rest of the layout. Within each hot zone, there are $4$ small cells (indicated by the red $\circ$'s) randomly dropped in order to meet the high traffic demands in the hot zone. Note that the model we use to drop the small cells and the users is exactly compliant with the parameters provided in the 3GPP standardization document \cite{3GPP}. The Macro/small cell powers and the pathloss models used in this experiment are identical to those used in Experiment 1. 
 
 We calculate the peak rates $R_{k,j}$ using the formulas (\ref{user-rates-massive-MIMO}) and (\ref{sinr-zfbf}) for ZFBF with pilot contamination. We assume that each Macro BS uses the same set  of $10$ pilots which are mutually orthogonal. Furthermore, we assume that the pilots used by the small cell BSs are orthogonal with the $10$ pilots used by the Macro BSs. Moreover, within each hot zone, $16$ mutually orthogonal pilots are used; $4$ pilots each for each of the small cell BSs in the hot zone. These $16$ pilots are then re-used in every hot zone of the layout.

\begin{figure}[t]
\centering
\includegraphics[scale = 0.5]{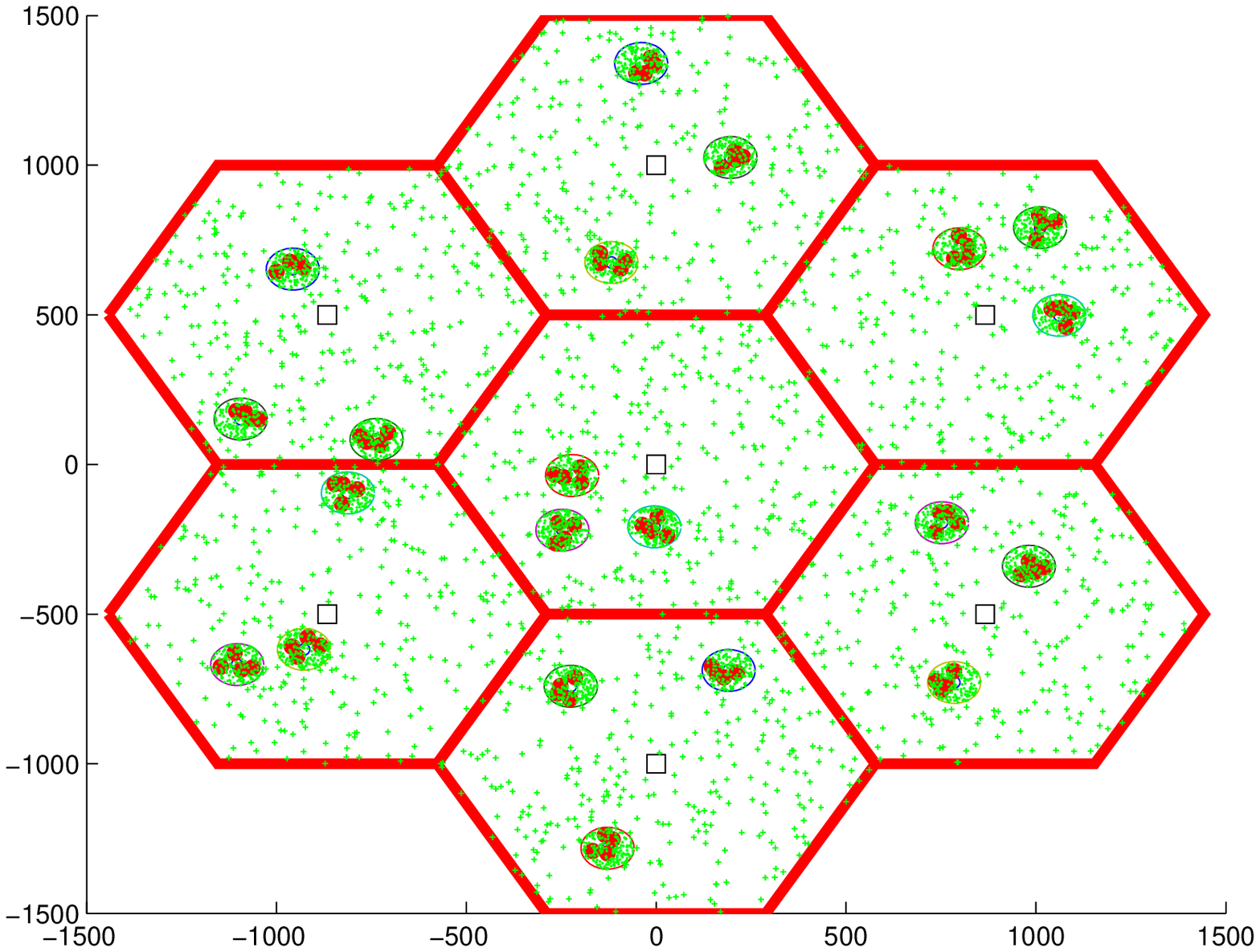}

\caption{A 3GPP HetNet scenario with small cell BSs deployed in hot zones}
\label{layout}
\end{figure}
As in Experiment 1, we run the distributed algorithm and the Max peak-rate scheme with $\gamma = 1$ (proportional fairness) for $100$ different realizations of the 3GPP layout of Fig.~\ref{layout} wherein the locations of the users, hot zones and the small cell BSs are generated randomly for each realization in compliance with 3GPP specifications while the locations of the Macro BSs remain fixed. Similar to Experiment 1, we compare the performance of the distributed algorithm and the Max peak-rate scheme by plotting the CDF of the throughput statistics. From Fig.~\ref{thruput-statistic}, we notice that the distributed algorithm provides $\geq 25\%$ gain over Max peak-rate in terms of the $5$ percentile throughput for about $50\%$ of the realizations. 

\begin{figure}
\centering
\includegraphics[scale=0.5]{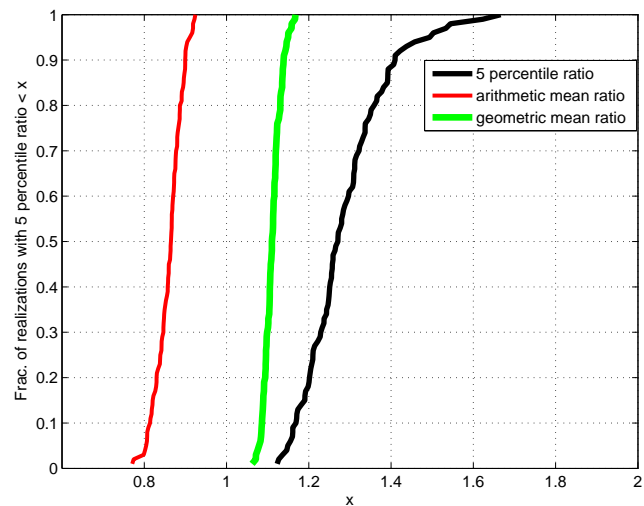}
\caption{Comparison of the proposed distributed algorithm and Max peak-rate in terms of the throughput statistics: $5$ percentile rate, arithmetic mean rate and geometric mean rate, for $\gamma = 1$ and the layout of Fig.~\ref{layout}.}
\label{thruput-statistic}
\end{figure}   

Finally, in Fig. ~\ref{bargraph}, we compare the performance of the proposed user-centric distributed algorithm with the Max peak-rate scheme for the layout in Fig.~\ref{layout} in terms of load balancing across various BSs. For every BS $j$, we first calculate the {\it load} which is the number of users $|\Kc_j|$ uniquely associated with BS $j$. Then, we plot in Fig.~\ref{bargraph} the per-BS load for the macro and the small cell BSs, where
within each set we sort the BSs in decreasing load order.  The superior performance of the proposed algorithm can be qualitatively 
appreciated from Fig.~\ref{bargraph}  by observing that the load achieved by our scheme is more evenly balanced, both across the two tiers, and within 
BSs of the same tier.

\begin{figure}
\centering
\includegraphics[scale=0.45]{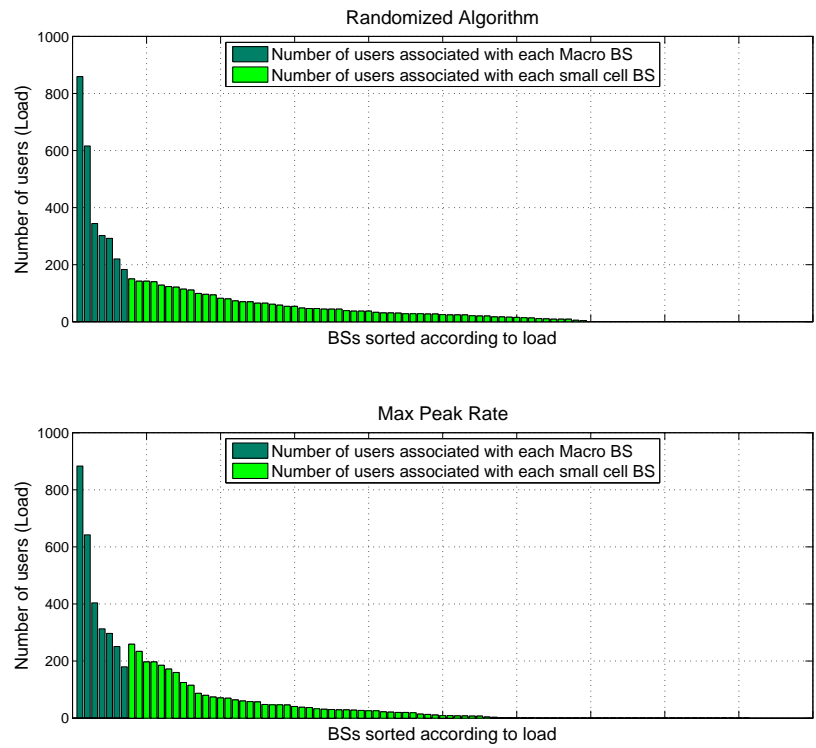}
\caption{Load distribution: proposed distributed algorithm vs. Max peak-rate association.}
\label{bargraph}
\end{figure}

\clearpage
\appendices

\section{Massive MIMO User Rates}  \label{massiveMIMO}

While the results and the schemes presented in this paper hold for any network characterized by a
set of user instantaneous rates $\{R_{k,j}\}$ and BS spatial multiplexing constraints $\{S_j\}$, 
it is worthwhile to specifically connect our treatment to massive MIMO performance analysis. 
The rate formulas presented here can be obtained, albeit at the cost of some effort, 
by particularizing the results found in several papers (in particular, see \cite{marzetta2010noncooperative,hoydis2011massive,Huh11}). 
For the sake of completeness, we restate some massive MIMO rate analysis results  in a unified notation consistent with this paper.

One use of the (complex discrete-time baseband) channel observed at the $k$-th user receiving antenna can be represented as
\begin{equation} \label{revsuca1}
y_{k,j} = \sum_{j \in \Jc} \sqrt{g_{k,j}} \hv_{k,j}^\herm \xv_j + z_{k},  
\end{equation}
where $g_{k,j} \in \RR_+$ denotes the large-scale channel power-gain coefficient between user 
$k$ and BS $j$, comprising distance-dependent pathloss and shadowing, $\xv_j \in \CC^{M_j}$ is the transmit signal vector of BS $j$, 
$z_k \sim \Cc\Nc(0,N_0)$ is the additive Gaussian noise sample at receiver $k$, and $\hv_{k,j}$ is the $M_j$-dimensional channel vector, formed 
by the small-scale fading coefficients. We assume i.i.d. Rayleigh fading, such that $\hv_{k,j}$ has i.i.d. elements $\sim \Cc\Nc(0,1)$. 
The transmitted signals are constrained by $\EE[\|\xv_{j}\|^2] \leq P_j$, where $P_j$ denotes the transmit power of BS $j$.
BS $j$ sends $S_j$ downlink data streams in each scheduling resource block. We let $\nu_j = S_j/M_j$ denote the spatial load
(number of downlink streams per BS antenna). 
With linear multiuser MIMO precoding,  each base station serving users $\Sc_j \subset \Kc$ with $|\Sc_j| = S_j$ forms its 
transmitted signal as $\xv_j = \sum_{k \in \Sc_j} \vv_{k,j} d_{k}$, where $\{d_{k}\}$ are mutually uncorrelated zero-mean 
data symbols with the same per-symbol average energy for all $k \in \Sc_j$ (we assume equal power per stream at each BS).

The precoding vectors $\{\vv_{k,j} : k \in \Sc_k\}$ are computed by BS $j$ as a function of its {\em Channel State Information} (CSI). 
We follow the CSI estimation scheme based on TDD with uplink-downlink (uplink-downlink) reciprocity 
as in  \cite{marzetta2010noncooperative} adapted to the heterogeneous network  at hand. 
Assuming block-fading, constant over time-frequency coherence blocks of $T$ channel uses, and letting 
$\max\{S_j\} \leq Q \leq T$ denote the uplink pilot dimension, the CSI is obtained by letting the active users in each cell send their uplink pilot signals
on the first $Q$ symbols on each slot. Then, downlink data transmission takes place in the remaining $T - Q$ symbols. 
We index the set of mutually orthogonal pilot signals by the set $\Qc = \{1, \ldots, Q\}$.  
Pilot signals are distributed across the BSs such that BS $j$ is given a subset 
$\Qc^{(j)} \subseteq \Qc$ of size $|\Qc^{(j)}| = S_j$ of mutually orthogonal pilots. 
We denote by $q(k)$ the pilot index of user $k$. In particular, if $k \in \Sc_j$, then $q(k) \in \Qc^{(j)}$. 
Also, we let $\Jc^{(q)} \subseteq \Jc$ denote the set of BSs which make use of pilot signal $q$, i.e., 
$\Jc^{(q)} = \{ j \in \Jc : q \in \Qc^{(j)}\}$.  
The pilot signal allocation, defined equivalently by the ensembles of sets $\{\Qc^{(j)} : j \in \Jc\}$ or 
$\{\Jc^{(q)} : q \in \Qc\}$, is optimized in some suitable way depending on  the topology of the network  
(see for example \cite{Huh11} for a thorough  analysis of optimized  pilot reuse schemes).  Here, we provide formulas that hold for 
any pilot allocation of the type considered in \cite{marzetta2010noncooperative,hoydis2011massive,Huh11}, 
i.e., where each cell is given a set of mutually orthogonal pilot signals, and sets of different cells may have non-empty intersection
(leading to pilot contamination). The specific optimization of the pilot allocation across base stations, in a multicell scenario, 
is well beyond the scope of this paper.   

The uplink signal block received at BS $j$ during the uplink training phase is given by 
\begin{equation} \label{ziofanale11}
\Ym_j^{\rm ul} = \sum_{\ell \in \Jc} \sum_{k \in \Sc_\ell} \sqrt{g_{k,j}} \hv_{k,j} \xiv_{q(k)}^\herm + \Zm^{\rm ul}_j, 
\end{equation}
where  $\xiv_{q} \in \CC^Q$ is the $q$-th pilot signal and $\Zm^{\rm ul}_j \in \CC^{M_j\times Q}$ with 
i.i.d. elements $\sim \Cc\Nc(0,N_0)$. We follow the CSI estimation approach given in 
\cite{marzetta2010noncooperative}, where BS $j$ obtains the estimate  of the downlink channel vector for user $k \in \Sc_j$ up to a real positive 
scaling factor and some bias additive terms known as {\em pilot contamination} by projecting
$\Ym_j^{\rm ul}$ along the pilot signal vector $\xiv_{q(k)}$.  Pilot contamination is due to the fact that, since $T$ is limited by the channel coherence time and bandwidth, 
then $Q$ cannot be arbitrarily large.  Hence, the $Q$ mutually orthogonal pilot signals must be reused 
by several BSs. In particular, the CSI estimate for user $k \in \Sc_j$, given by\footnote{We let
$\widetilde{\zv}_{k,j}^{\rm ul} =  \Zm_j^{\rm ul} \xiv_{q(k)} \|\xiv_{q(k)} \|^{-2}$ denote the projected noise vector
with i.i.d. components $\sim \Cc\Nc(0,\sigma^2)$, where $\sigma^2 \eqdef  \frac{N_0}{Q P_u}$ and where 
$P_u$ denotes the energy per symbol of the uplink pilot signals.}
\begin{equation} \label{hestimate}
\widehat{\hv}_{k,j} = \frac{\Ym_j^{\rm ul} \xiv_{q(k)}}{\|\xiv_{q(k)} \|^2} 
= \sum_{k' : q(k') = q(k)} \sqrt{g_{k',j}} \hv_{k',j} + \widetilde{\zv}_{k,j}^{\rm ul}, 
\end{equation}
contains the linear combination of the channels from all users $k'$ using the same pilot signal $q(k)$ (where $k \in \Sc_j$ and the other $k' \neq k$ are
active in other cells) to BS $j$. 

The most popular and simplest multiuser MIMO downlink precoding methods, widely analyzed and also implemented in practice,  
are conjugate beamforming (CBF) \cite{marzetta2010noncooperative} and zero-forcing beamforming (ZFBF) \cite{hoydis2011massive,Huh11}.

{\bf User SINR with CBF:} 
Particularizing the analysis in \cite[Th. 1]{Huh11} to the notation and CSI estimation given above, 
we find that the Signal to Interference plus Noise Ratio (SINR) at user $k$ receiver served
by BS $j$ under the system assumptions given above, for large $M_j$ and $\nu_j = S_j/M_j$, 
is closely approximated (in the sense of Lemma \ref{massive-lemma}) by the {\em deterministic} quantity:
\begin{equation} \label{sinr-cbf}
\SINR_{k,j} =  \frac{g^2_{k,j} \SNR_j / \nu_j}{{\displaystyle \eta + \sum_{\ell \in \Jc} g_{k,\ell} \SNR_\ell 
+ \sum_{\ell \in \Jc^{(q(k))} :\ell \neq j}
g^2_{k,\ell}  \SNR_\ell / \nu_\ell}},
\end{equation}
where $\SNR_j \eqdef P_j/N_0$, and where $\eta \geq 1$ is a normalization factor, common to all BSs, chosen to ensure that the transmit power 
constraint is not violated by any BS \cite{marzetta2010noncooperative}. In practice, $\eta$ can be adjusted adaptively by measuring 
the transmit power at each BS (which is a function of the precoding vectors and, as a consequence, of the estimated channel vectors 
(\ref{hestimate})) averaged over a suitably chosen time window. Notice also that in the limit of infinite antennas and finite number of downlink streams, 
i.e., $\nu_j \rightarrow 0$ for all $j$, the choice of $\eta$ becomes irrelevant, and the SINR converges to
the well-known massive MIMO expression $\SINR_{k,j} = \frac{g^2_{k,j}}{\sum_{\ell \in \Jc^{(q(k))} :\ell \neq j}
g^2_{k,\ell}}$, in the symmetric case where $\SNR_j$ and $\nu_j$ are identical for all BS, 
as derived in  \cite{marzetta2010noncooperative}. 

{\bf User SINR with ZFBF:} 
Particularizing the analysis in \cite[Th. 2]{Huh11}, we find that the Signal to Interference plus Noise Ratio (SINR) at user $k$ receiver served
by BS $j$ under the system assumptions given above, for large $M_j$ and $\nu_j = S_j/M_j$, is closely approximated (in the sense of
Lemma \ref{massive-lemma}) by the {\em deterministic} quantity:
\begin{equation} \label{sinr-zfbf}
\SINR_{k,j} = \frac{(1 - \nu_j) g^2_{k,j} \SNR_j/\nu_j}{{\displaystyle \eta + \sigma^2 g_{k,j} \SNR_j + \sum_{\ell \in \Jc : \ell \neq j} g_{k,\ell} \SNR_\ell 
+ \sum_{\ell \in \Jc^{(q(k))} :\ell \neq j} (1 - \nu_\ell) g^2_{k,\ell} \SNR_\ell / \nu_\ell }}. 
\end{equation}
Comparing (\ref{sinr-zfbf}) with (\ref{sinr-cbf}) we notice that the effect of ZFBF consists of 
decreasing the beamforming gain and the pilot contamination effect by the quantity $(1 - \nu_j)$, 
due to zero-forcing precoding,  and reducing the intra-cell interference from $g_{k,j} \SNR_j$ to $\sigma^2 g_{k,j} \SNR_j$. 
In fact, in the case of ideal channel estimation (i.e., $\sigma^2  = 0$), the intra-cell interference with ZFBF  is exactly zero. 
Notice also that, in the limit of $\nu_j \rightarrow 0$ for all $j$,  the SINRs with ZFBF and CBF coincide, 
confirming the fact that CBF and ZFBF are equivalent in the regime of very large antennas and finite 
number of  users per  BS \cite{marzetta2010noncooperative}. 

{\bf User instantaneous rates:}
At this point, assuming Gaussian codebooks and taking into account that the downlink data transmission phase takes place on
$T - Q > 0$ dimensions, for each slot of $T$ dimensions, the user rates expressed in bit/dimension are given by 
\begin{equation} \label{user-rates-massive-MIMO}
R_{k,j} = (1 - Q/T) \log_2 \left ( 1 + \SINR_{k,j} \right ).
\end{equation}

\section{Proof of Theorem \ref{th-feasibility}} \label{proof:feasibility}

We represent the network by a bipartite  graph $\Gc = (\Jc, \Kc, \Ec)$ where $\Jc$ is the set of BS nodes, 
$\Kc$ is the set of user nodes, and $\Ec = \Jc \times \Kc$ is the set of edges indicating possible association 
(for simplicity, here we let $\Jc_k = \Jc~\forall~k$). An integer scheduling configuration corresponds to a collection of 
edges $\Fc \subseteq \Ec$, such that each BS $j \in \Jc$ is incident to at most $S_j$ edges in 
$\Fc$, while each user $k$ is incident to at most one edge in $\Fc$. 
When $S_j = 1~\forall~j \in \Jc$, an integer scheduling configuration $\Fc$ corresponds to a {\it matching} 
in $\Gc$. For $S_j > 1$, we can think of an integer scheduling configuration $\Fc$ as a {\em generalized matching}.
We now associate a point in $\mathbb{R}^{|\Ec|}$ to every integer scheduling configuration $\Fc$. 
For this purpose, given an integer scheduling configuration $\Fc$, let its {\it incidence vector} be $\sigma$ where 
$\sigma_{k,j} = 1$ if $(k,j) \in \Fc$ and $0$ otherwise. Let $\Omega$ denote the set of incidence vectors 
where each incidence vector corresponds to an integer scheduling configuration.
By time-sharing among such integer scheduling configurations, any {\it feasible association configuration} 
in the convex hull of $\Omega$ can be achieved in the sense of long-term time average. Let 
\begin{align}
P^{\prime} = \textrm{coh} (\Omega)
\end{align}
denote the convex polytope obtained by taking the convex hull of the points in $\Omega$. 
Also, let $P$ denote the convex polytope corresponding to the set of linear constraints (\ref{matchingconstJ})--(\ref{ineq}), i.e., 
containing all the feasible association configurations. 
The relation between the convex polytopes $P$ and $P^{\prime} $ is not clear a priori. If one could show that $P = P^{\prime}$, 
then any feasible association configuration can be realized  by first expressing the vector of user activity fractions 
as a convex combination of integer scheduling configurations  in $\Omega$, and then 
time sharing the transmission slots among those configurations with scheduling dictated by the 
convex combination (see the observation made before Theorem \ref{th-feasibility} in Section \ref{sec:system-optimization}). 
Hence, proving $P = P^{\prime}$ implies the proof of Theorem \ref{th-feasibility}.
We shall prove this assertion by showing that both the 
relations $P \subseteq P^{\prime}$ and $P^{\prime} \subseteq P$ hold.

\begin{proposition}
$P^{\prime} \subseteq P$.
\label{direct}
\end{proposition}

\begin{IEEEproof}
Consider any integer scheduling configuration $\sigma \in \Omega$. 
It is easy to check that $\sigma$ satisfies the constraints (\ref{matchingconstJ})--(\ref{ineq}). 
Thus, $\Omega \subseteq P$ holds and since $P$ is a convex polytope, 
$P^{\prime} = \textrm{coh}(\Omega)$ is also a subset of $P$. 
\end{IEEEproof}

\begin{proposition}
$P \subseteq P^{\prime}$.\label{converse}
\end{proposition}
We state a series of lemmas which provide a proof of Proposition \ref{converse}. 
While we give proofs for certain lemmas, the other lemmas are well known and the reader is referred to the relevant literature in combinatorial optimization (see \cite{vondraknotes, schrijverbook} for example). The goal is to show that the set of extreme points (vertices) of $P$ is included in the set of incidence vectors of integer scheduling configurations $\Omega$. Once this is shown, we have our result since  $P = \textrm{coh}(ext(P)) \subseteq \textrm{coh} (\Omega) = P^{\prime}$ 
where $ext(P)$ is the set of the extreme points of $P$. We re-write $P$, i.e., the constraints (\ref{matchingconstJ})--(\ref{ineq}) as:
\begin{align}
P = \left \{\alphav: \begin{bmatrix*}[r]\Jm \\ \Km \\ -{\mathbf I} \end{bmatrix*}\alphav \leq \begin{bmatrix*}[r] \sv \\ {\mathbf 1} \\ {\mathbf 0}\end{bmatrix*} \right \}
 = \{\alphav: \Am\alphav \leq \bv\}, \label{standard-form}
\end{align} 
where {\footnotesize $\Am =  \begin{bmatrix*}[r]\Jm \\ \Km \\ -{\mathbf I} \end{bmatrix*}$} and {\footnotesize $\bv = \begin{bmatrix*}[r] \sv \\ {\mathbf 1} \\ {\mathbf 0}\end{bmatrix*}$}. 
Here, $\Jm$ is a matrix of dimensions $|\Jc| \times |\Ec|$ with elements in the binary set $\{0,1\}$, 
where columns are indexed by edges in $\Ec$ and the rows are indexed by the BSs in $\Jc$. 
Each column has exactly one $1$ corresponding to the BS on which the edge is incident. 
$\sv $ is a $|\Jc| \times 1$ column vector with elements $S_j~\forall~j \in \Jc$. 
Similarly, $\Km$ is a matrix of dimensions $|\Kc| \times |\Ec|$ with elements in $\{0,1\}$, where columns are again indexed by the edges in $\Ec$ and the rows are indexed by the users in $\Kc$. 
Again, each column has exactly one $1$, corresponding to the user on which the edge is incident. 
${\mathbf 1}$ is a $|\Kc| \times 1$ all-1 column vector. 
${\mathbf I}$ is the $|\Ec| \times |\Ec|$ identity matrix. 
Note that $\Gm = \begin{bmatrix*}[r] \Jm \\ \Km \end{bmatrix*}$ is the 
{\it incidence matrix} of the bipartite graph $\Gc$, i.e., $\Gm$ is a matrix of dimensions $(|\Jc|+|\Kc|) \times |\Ec|$ and elements in $\{0,1\}$, 
where columns are indexed by edges in $\Ec$ and each column has exactly two $1$'s, corresponding to the two vertices of 
the edge (one vertex in $\Jc$ and the other in $\Kc$). 

\begin{definition}
{\bf Extreme Point}: A point $\vv$ in $P$ is said to be an extreme point  of $P$ if it cannot be expressed as a convex combination of points in $P \setminus \{\vv\}$.
\end{definition}
Let $\vv$ be an extreme point of the convex polytope $P$ given by (\ref{standard-form}). Then,  $\vv$ must satisfy the following lemmas: 
\begin{lemma}
 There are $|\Ec|$ constraints in $\Am\alphav \leq \bv$ which are tight at $\vv$, i.e., $\av_i^\transp\vv = b_i~\forall~i \in \{1, \ldots, |\Ec| \}$ and in addition, $\av_1, \ldots, \av_{|\Ec|}$ are linearly independent.
 \label{tight}
 \end{lemma}
\begin{IEEEproof}
Consider $T = \{ \av_j: \av_j^\transp \vv = b_j\}$. If $\textrm{dim}(span(T)) < |\Ec|$, then there exists $\dv \neq 0$ such that $\dv$ is orthogonal to $span(T)$, i.e., for all $\av_j \in T$, $\av_j^\transp \dv = 0$ and therefore $\av_j^\transp(\vv \pm \epsilon \dv) = \av_j^\transp \vv = b_j$. For all other constraints, $\vv$ satisfies strict inequality, i.e., $\av_i^\transp \vv < b_i$, so there is some sufficiently small $\epsilon >0$ such that $\av_i^\transp(\vv+\epsilon \dv) \leq b_i$ and $\av_i^\transp(\vv-\epsilon \dv) \leq b_i$. This means that $\vv + \epsilon \dv$ and $\vv - \epsilon \dv$ are in $P$ which in turn implies that $\vv = \frac{1}{2}(\vv + \epsilon \dv) + \frac{1}{2}(\vv - \epsilon \dv)$ is expressed as a convex combination of two other feasible points. This contradicts the fact that $\vv$ is an extreme point.
\end{IEEEproof}
\begin{lemma}
 $\vv$ is the unique solution to the $|\Ec|$ constraints which are tight from Lemma \ref{tight}. 
 \label{unique}
 \end{lemma}
 \begin{IEEEproof}
The set of $|\Ec|$ linear equations from Lemma \ref{tight} is a rank $|\Ec|$ system of linear equations in $|\Ec|$ dimensions. Thus, $\vv$ is the unique solution to the system.
\end{IEEEproof}
It follows that every extreme point (or vertex) of $P$ is a unique solution to the linear system obtained from the tightness of $|\Ec|$ constraints in the set of constraints $\Am \alphav \leq \bv$.
\begin{definition}
{\bf Totally Unimodular Matrix}: A matrix $\Gm$ is said to be {\it totally unimodular} if every square submatrix of $\Gm$ has determinant $0$, $+1$ or $-1$.  
\end{definition}
\begin{lemma}
For all bipartite graphs $\Gc$, the incidence matrix $\Gm$ is totally unimodular.
\label{tum}
\end{lemma}
\begin{lemma}
If $\Gm$ is totally unimodular, then $\begin{bmatrix*}[r] \Gm \\ -\mathbf{I} \end{bmatrix*}$ is totally unimodular.
\label{tum-i}
\end{lemma}
See \cite{vondraknotes} for proofs of Lemmas \ref{tum} and \ref{tum-i}.
In particular, the incidence matrix $\Gm = \begin{bmatrix*}[r] \Jm \\ \Km \end{bmatrix*}$ of the bipartite graph $\Gc$ is totally unimodular from Lemma \ref{tum}. 

Let $\vv$ be a vertex of $P$. From Lemmas \ref{tight} and \ref{unique}, there exists a rank $|\Ec|$ square submatrix $\Am^{\prime}$ of $\Am$ such that $\Am^{\prime}\vv = \bv^{\prime}$ and $\vv$ is the unique solution to the system. 
\begin{lemma}
$\vv$ is an integer vector and is in the set of integer scheduling configurations $\Omega$.
\end{lemma}
\begin{IEEEproof}
 $\Am^{\prime}$ is a full rank square submatrix of $\begin{bmatrix*}[r] \Gm \\ -\mathbf{I} \end{bmatrix*}$ and since $\begin{bmatrix*}[r] \Gm \\ -\mathbf{I} \end{bmatrix*}$ is totally unimodular from Lemma \ref{tum-i}, we have that $\det~\Am^{\prime} = \pm 1$.
Now by Cramer's rule, we have the $i$-th component $v_i$ of $\vv$ as:
\begin{align}
v_i = \frac{\det (\Am^{\prime}_i | \bv^{\prime})}{\det (\Am^{\prime})}
\end{align}
where $\Am^{\prime}_i | \bv^{\prime}$ is $\Am^{\prime}$ with the $i$-th column replaced by $\bv^{\prime}$. Note that $\bv$
has all integer elements, implying that $\bv^{\prime}$ is an integer vector. Thus, with $\bv^{\prime}$ being an integer vector and $\det(\Am^{\prime}) = \pm 1$, we conclude that $v_i$ is an integer. 
Now, given that $\vv$ is an integer vector and it satisfies (\ref{matchingconstJ})--(\ref{ineq}), 
the only possible way for which this can happen is that $\vv$ is an integer scheduling configuration,
i.e., $\vv \in \Omega$. This concludes the proof of Proposition $2$.
\end{IEEEproof}

\section{Proof of Theorem \ref{BS-NUM}} \label{proof-BS-NUM}

The Lagrangian corresponding to (\ref{NUM-local}) is
\begin{equation} \label{lagrangian-local}
\widecheck{L}(\alphav_{j},\mu) = \sum_{k \in \Kc_j}\frac{( \alpha_{k,j} R_{k,j}) ^{1-\gamma}}{1-\gamma} - \mu(\sum_{k \in \Kc_j} \alpha_{k,j} - S_j)
\end{equation}
where $\alphav_j = (\alpha_{1,j}, \ldots, \alpha_{|\Kc_j|,j})$ and $\mu \geq 0$. 
Since we assume $\gamma \geq 1$, the optimal $\alphav_j$ must have strictly positive components. 
(see Remark \ref{remark-gamma1}). 
By taking the partial derivative of (\ref{lagrangian-local}) with respect to $\alpha_{k,j}$, we obtain the necessary and sufficient KKT conditions 
for optimality in the form 
\begin{equation}
\alpha_{k,j} \leq \frac{R_{k,j}^{\rho-1}}{\mu^\rho}
\label{KKT-local-proof}
\end{equation}
where (\ref{KKT-local-proof}) must hold with equality for the variables $\alpha_{k,j}$ which are strictly less than $1$ at the optimal solution. In addition, (\ref{matchingconstJ-local}) must hold with equality since all resources are exhausted at the optimal solution, i.e.,
\begin{equation}\label{exhaust}
\sum_{k \in \Kc_j} \alpha_{k,j} = S_j
\end{equation}
Using the ordering (\ref{sort-users}) in (\ref{KKT-local-proof}),  we obtain an explicit expression of the optimal $\alphav_j$
in terms of the Lagrangian multiplier $\mu$ as
\begin{equation} \label{alpha-interm}
\alpha_{k,j} = \left \{ \begin{array}{ll}
1, & \;\;\; \mbox{for} \;\; 1\leq k \leq k^*-1 \\
\frac{R_{k,j}^{\rho-1}}{\mu^\rho} & \;\;\; \mbox{for} \;\; k^* \leq k \leq |\Kc_j| \end{array} \right .
\end{equation}
where  $k^* \in \{1, \ldots, |\Kc_j|\}$ is such that $R_{k^*-1,j}^{\rho-1} \geq \mu^\rho  > R_{k^*,j}^{\rho-1}$. 
Substituting (\ref{alpha-interm}) in (\ref{exhaust}), we can solve for $\mu$ and get
\begin{equation}\label{eta-calc}
\mu^{\rho}  =  \frac{\sum \limits_{k=k^*}^{|\Kc_j|}R_{k,j}^{\rho-1}}{S_j-k^*+1}.
\end{equation}
Substituting (\ref{eta-calc}) in (\ref{alpha-interm}), we finally obtain (\ref{alpha-final}).

By the sufficiency of the KKT conditions, the value of $k^*$ can be found as follows: the condition $R_{k^*-1,j}^{\rho-1} \geq \mu^\rho > R_{k^*,j}^{\rho-1} $ 
with $\mu^\rho$ given by (\ref{eta-calc}) is sequentially tested for tentative values of $k^* = 1,2,3,\ldots$ and the search is stopped (and the corresponding $k^*$ is chosen) 
as soon as this condition is satisfied.

\bibliographystyle{IEEEtran}
\bibliography{dilip-ref}

\end{document}

%% file: macros.tex
\long\def\comment#1{}

\newfont{\bbb}{msbm10 scaled 700}

\newfont{\bb}{msbm10 scaled 1100}
\newcommand{\CC}{\mbox{\bb C}}

\newcommand{\RR}{\mbox{\bb R}}

\newcommand{\EE}{\mbox{\bb E}}

\newcommand{\av}{{\bf a}}
\newcommand{\bv}{{\bf b}}

\newcommand{\dv}{{\bf d}}

\newcommand{\hv}{{\bf h}}

\newcommand{\jv}{{\bf j}}

\newcommand{\pv}{{\bf p}}

\newcommand{\rv}{{\bf r}}
\newcommand{\sv}{{\bf s}}

\newcommand{\vv}{{\bf v}}
\newcommand{\xv}{{\bf x}}

\newcommand{\zv}{{\bf z}}

\newcommand{\Am}{{\bf A}}

\newcommand{\Gm}{{\bf G}}

\newcommand{\Jm}{{\bf J}}
\newcommand{\Km}{{\bf K}}

\newcommand{\Ym}{{\bf Y}}
\newcommand{\Zm}{{\bf Z}}

\newcommand{\Ac}{{\cal A}}

\newcommand{\Cc}{{\cal C}}

\newcommand{\Ec}{{\cal E}}
\newcommand{\Fc}{{\cal F}}
\newcommand{\Gc}{{\cal G}}

\newcommand{\Jc}{{\cal J}}
\newcommand{\Kc}{{\cal K}}

\newcommand{\Nc}{{\cal N}}

\newcommand{\Qc}{{\cal Q}}

\newcommand{\Sc}{{\cal S}}

\newcommand{\alphav}{\hbox{\boldmath$\alpha$}}
\newcommand{\betav}{\hbox{\boldmath$\beta$}}

\newcommand{\lambdav}{\hbox{\boldmath$\lambda$}}

\newcommand{\xiv}{\hbox{\boldmath$\xi$}}

\renewcommand{\det}{{\hbox{det}}}

\newcommand{\SINR}{{\sf SINR}}
\newcommand{\SNR}{{\sf SNR}}

\newcommand{\eqdef}{\stackrel{\Delta}{=}}

\newcommand{\herm}{{\sf H}}

\newcommand{\transp}{{\sf T}}

